\documentclass{article}
\usepackage{graphicx}  
\usepackage{amsmath}   
\usepackage[noadjust]{cite}
\usepackage{amssymb}   
\usepackage{bm} 
\usepackage{dcolumn}
\usepackage{color}
\usepackage{mathrsfs}
\usepackage{amsfonts}
\usepackage{varioref}
\usepackage{braket}

\RequirePackage[colorlinks,citecolor=blue,urlcolor=magenta,linkcolor=blue]{hyperref}

\def\PPEU{particle production in an expanding universe}

\def\TDHO{time dependent harmonic oscillator}
\labelformat{section}{Section #1} 
\labelformat{subsection}{Section #1} 
\labelformat{subsubsection}{Section #1}
\labelformat{subsubsubsection}{Section #1}
\labelformat{equation}{Eq.~(#1)} 
\labelformat{figure}{Fig.~#1} 
\labelformat{subfigure}{Fig.~\thefigure#1} 
\labelformat{table}{Tab.~#1} 
\labelformat{appendix}{Appendix #1}
\title{Generalized Schwinger effect and particle production in an expanding universe}

\author{Karthik Rajeev\footnote{karthik@iucaa.in}$~^{1}$,~ 
Sumanta Chakraborty\footnote{sumantac.physics@gmail.com}$~^{2}$~and~
T. Padmanabhan\footnote{paddy@iucaa.in}$~^{1}$\\
{\small{$^{1}$IUCAA, Post Bag 4, Ganeshkhind, Pune University Campus, Pune 411007, India}}\\
{\small{$^{2}$School of Physical Sciences, Indian Association for the Cultivation of Science, Kolkata-700032, India}}}

\begin{document}

\maketitle
\begin{abstract}

We discuss several aspects of particle production in: (a) time dependent electric field and (b) expanding Friedmann background. In the first part of the paper, we provide an algebraic  mapping between the differential equations describing these two phenomena. This mapping allows a direct comparison between (a) and (b) and we highlight several interesting features of both cases using this approach. We  determine the form of the (equivalent) electric field corresponding to different Friedmann spacetimes and discover, for example,  a time-dependent electric field which, in a specific limit, leads to a Planck's spectrum of particles. We also discuss the conditions under which the particle production in an expanding background will be non-analytic in the parameter which encodes the coupling to the curved spacetime, in close analogy with the generalized Schwinger effect.  In the second part of the paper, we  study the situation in which both time dependent electric field and an expanding background are simultaneously present.  We compute particle production rate in this context by several different methods paying special attention to its limiting forms and possible non-analytic behaviour. We also clarify several conceptual issues related to definitions of in-vacuum and out-vacuum in these systems.   

\end{abstract}

\section{Introduction}

Two examples of particle production in external backgrounds --- discussed extensively in the literature --- corresponds to: (a) quantum field theory of a complex scalar field in an external, homogeneous, electric field \cite{schwinger1951gauge} and (b) quantum field theory of a scalar field in an expanding Friedmann background \cite{parker1967creation}.  We will be dealing with several aspects of these two systems in this paper.

The first of these two examples include the famous Schwinger effect \cite{schwinger1951gauge,Heisenberg1936}, corresponding to a constant electric field, and its   generalizations which deal with time dependent electric fields (see e.g., Refs.\cite{time1,time2,time3,time4,time5,time6,time7,time8,time9,time10}. In the case of constant electric field, the number density of particles produced from the vacuum has a non-analytic dependence on the coupling constant, which implies that this result cannot be obtained from perturbative QED.  The situation changes when the electric field is time dependent (see for e.g., Refs. \cite{sauter2,sauter4,sauter5,sauter6,sauter7}).  Broadly speaking, if the electric field is sharply localized in a time interval, the particle production rate exhibits an analytic dependence on the coupling constant. It is possible to construct examples in which the particle production rate makes a smooth transition between non-analytic to analytic behaviour (with respect to the coupling constant), when a parameter which controls the time dependence of the electric field is varied. 

In the first part of this paper, we will show that there exists a \textit{purely algebraic} correspondence between the differential equations governing the scalar field in the cases (a) and (b) mentioned above. (Though some authors have noticed a parallel between particle production in an expanding universe and Schwinger effect in the past --- one of the earliest works we know being Ref. \cite{Padmanabhan:1991uk}  and a more recent one being \cite{Martin2007} ---  the utility of this parallel has not been adequately exploited in the literature.) Using this correspondence, it is possible to translate the results obtained in the case of a time dependent electric field to those in an expanding Friedmann background and vice-versa. For example, it turns out that the constant electric field case can be mapped to a radiation dominated universe while the de Sitter universe maps back to a singular electric field. Further, the Milne universe maps to an electric field in flat spacetime which produces a  Planckian spectrum of particles, in a specific limit, thereby providing yet another `black hole analogue' \cite{scrktp}.

Just as the particle production in an external electric field vanishes when a coupling constant in the problem goes to zero, the particle production in an expanding universe will vanish when the parameter describing  the expansion of the universe goes to zero, thereby reducing the Friedmann universe to a flat spacetime. (For example, the spatially flat de Sitter universe will become flat spacetime when $H\to 0$.) This vanishing of the particle production in the expanding universe can, again, have either an analytic or non-analytic dependence in the relevant coupling constant, just as it happens in the case of time-dependent electric field. In a previous work \cite{Rajeev2018},  we provided a  criterion  to distinguish between these two types of behaviour for a broad class of time dependent electric fields. The mapping between the two cases (a) and (b) allows us to translate the results in Ref. \cite{Rajeev2018} to the case of particle production in Friedmann universes and obtain a criterion for analytic versus non-analytic dependence of the coupling constant in this context. 

In the second part of the paper we study situations in which both time dependent electric field as well as background expansion are present in the form of a time dependent electric field in the de Sitter universe. (The effects of pair production in such settings are believed to be relevant in the study of inflationary magnetogenesis; see e.g., Refs. \cite{magnetogenesis,magnetogenesis2,Chakraborty:2018dmj}). The particle production in this context should reduce to two previously known limits when we switch off the electric field or the de Sitter expansion. To understand these limits properly, we first describe certain peculiar features which arise in the study of particle production in the de Sitter background. The conventional method for studying the particle production in an expanding background is based on calculating the Bogoliubov coefficients between the in-modes and the out-modes. The straightforward application of this method works in a de Sitter background only when $(M/H) > 3/2$ (where $M$ is the mass of the scalar field quanta and $H$ is the Hubble constant) and fails when $(M/H) < 3/2$. We provide a careful discussion of this failure and show how particle production can be computed for all values of $(M/H)$ by using a different method \cite{Mahajan:2007qc,Mahajan2008}. This method uses the Schr\"{o}dinger picture description and evolves an asymptotic in-vacuum state wave function into the future using the Schr\"{o}dinger equation. Expanding this wave function in terms of instantaneous mode functions one can describe particle production in de Sitter for all values of $M/H$. It turns out that similar issues arise when we study particle production due to the combined effect of electric field and de Sitter expansion and can be satisfactorily addressed by using a similar procedure. We also pay careful attention to the two limits $E\to 0$ and $H\to 0$ and ensure that our results have the correct limiting forms. (This has not been the case in some of the previous literature in this subject.)

The structure of the paper is as follows: In \ref{sec1}, we show that there is a well defined algebraic correspondence between the differential equations describing the pair production of a massive scalar field in Friedmann universe and a massive complex scalar in a homogeneous but time dependent electric field in Minkowski spacetime. We illustrate this correspondence in physically relevant examples and their useful limiting cases. Using the correspondence established in \ref{sec1}, and the techniques developed in Ref. \cite{Rajeev2018} to study generalized Schwinger effect, we analyze the analytic versus non-analytic dependence  of pair production in Friedmann universe in \ref{section2}. In \ref{puredS}, we briefly review pair creation in de Sitter spacetime using various methods. We review the standard procedure based on  mode functions (which works best for $M/H>3/2$) and an alternative procedure, based on the  instantaneous diagonalization of the Hamiltonian (which works for all values of $M/H$).  Finally, we discuss the Schwinger effect in de-Sitter in \ref{dSplusconstE} using two approximate methods (namely the Landau procedure and the Euclidean action method) and one exact method (using the mode functions) and compare the results. In \ref{sec2}, we complement the results of \ref{sec1}, by studying the pair production in a homogeneous but time dependent electric field in de-Sitter universe. 

The last section summarizes the paper and gives a list of new --- conceptual and technical --- results obtained in this work.
\section{Correspondence between generalized Schwinger effect and pair production in an expanding universe}\label{sec1}

Two time dependent quantum systems of  importance, in the discussion  of quantum fields in non-trivial backgrounds, are: (a) a massive scalar field in the Friedmann universe and (b) a massive complex scalar field in a homogeneous but time dependent electric field. In this section we will briefly review the algebraic correspondence between these two and study several properties using this correspondence.  One of the common features of these systems is that the time dependence of the background fields (viz., the metric and the electric field), leads to the creation of particle pairs from the vacuum. Though,  these effects are usually studied separately in literature for  these two systems, it is possible to establish an algebraic mapping between them. This allows us to translate the particle production rate in one case to the same in the other. In this section, we will establish this precise correspondence (see e.g., \cite{Padmanabhan:1991uk}) and demonstrate possible applications.

The Friedmann metric describing a spatially flat, isotropic and homogeneous universe contains a single time dependent parameter $a(t)$, and is given by:
\begin{align}\label{defineFriedmann}
ds^{2}=-dt^2+a^{2}(t)|d\mathbf{x}|^2~.
\end{align}
 It is convenient to transform  to the conformal time coordinate $\eta$ defined by $d\eta=dt/a(t)$, such that the line element takes the following form, 
\begin{align}\label{frwmetric}
ds^2=a(\eta)^2\left(-d\eta^2+|d\mathbf{x}|^2\right)~,
\end{align}
which is conformally flat in the $\eta$ coordinate. We will consider a conformally coupled, massive, scalar field in this background with the action:
\begin{align}\label{conformaction}
\mathcal{A}=\frac{1}{2}\int d^{4}x\sqrt{-g}\phi\left[\Box-\frac{R}{6}-M^2\right]\phi~.
\end{align}
In the massless limit, presence of the Ricci scalar term ensures that this action is invariant under conformal transformations. Thus, for the conformally flat background given in \ref{freqFriedmann}, when $M=0$, the above action is equivalent to that of a massless field $\Phi=a\phi$, in flat spacetime. However, when $M\neq 0$, the action in \ref{conformaction} transforms to the form
\begin{align}
\mathcal{A}=\frac{1}{2}\int d^{4}x\, \Phi\left[\Box_{\rm flat} -M^2a^2\right]\Phi~.
\end{align}
where, $\Box_{\rm flat}$ is the flat spacetime Laplacian and $\Phi=a\phi$. We will Fourier transform the field $\Phi$ in the spatial coordinates and introduce the Fourier modes $\Phi_{\mathbf{k}}$. In terms of $\Phi_{\mathbf{k}}$, the action simplifies to that of a bunch of time dependent harmonic oscillators, each labelled by the wavenumber $\mathbf{k}$. The corresponding Lagrangian associated with $\Phi_{\mathbf{k}}$ is given by:
\begin{align}
L_{\mathbf{k}}= \left[\frac{1}{2}|\Phi^{'}_{\bf k}|^{2}
-\frac{1}{2}\left(k^{2}+a^{2}M^{2}\right)|\Phi _{\bf k}|^{2}\right]~,
\end{align}
where, `prime' denotes derivative with respect to the conformal time coordinate $\eta$. Variation of the above Lagrangian with respect to $\Phi_{\mathbf{k}}$, yields its equation of motion, which can be written as,
\begin{align}\label{diffforfrw}
\Phi_{\mathbf{k}}''+\left\{k^2+M^2a^2(\eta)\right\}\Phi_{\mathbf{k}}=0~,
\end{align}
with $k^2=|\mathbf{k}|^2$. This describes a time dependent harmonic oscillator of unit mass and frequency $\omega_{\mathbf{k}}(\eta)$ given by 
\begin{align}\label{freqFriedmann}
\omega^2_{\mathbf{k}}(\eta)=k^2+M^2a(\eta)^2
\end{align}

We will next describe the correspondence between this differential equation and the one which arises in the case of a time dependent electric field in flat spacetime.  To do this, we will consider a complex scalar field $\Psi$ in the background of a homogeneous, but time dependent, electric field in flat spacetime (with the metric $ds^2=-d\eta^2+d\mathbf{x}^2$) and obtain an algebraic one-to-one correspondence between the two systems. Without loss of generality, let us assume that the electric field is along the $z$ direction and choose the vector potential to be $A_a=\{0,0,0,A_z(\eta)\}$. The field equation for the complex scalar field, minimally coupled to this background electric field, is given by
\begin{align}
\left(\partial_{a}-iqA_a\right)\left( \partial^{a}-iqA^a\right)\Psi-m^2\Psi=0.
\end{align} 
As in the previous case, we introduce the Fourier transform $\Psi_{\mathbf{p}}$ of the complex scalar field $\Psi$, defined by 
\begin{align}
	\Psi(\mathbf{x},\eta)=\int_{-\infty}^{\infty} \frac{dp_z}{2\pi}\int \frac{d^2\mathbf{p}_{\perp}}{(2\pi)^2}e^{i\mathbf{p}.\mathbf{x}}\Psi_{\bf p}(\eta),
\end{align}
to describe the dynamics. In the above expression, we have separated the momentum integral into that of the longitudinal ($p_z$) and the transverse ($\mathbf{p}_{\perp}\equiv \mathbf{p}-p_z\hat{\mathbf{z}}$) components for later convenience. The equation of motion satisfied by $\Psi_{\mathbf{p}}$ takes the form
\begin{align}\label{eomcomplex}
\Psi_\mathbf{p}''+\left\{m^2+p_{\perp}^2+(p_z+qA_z)^2\right\}\Psi_{\mathbf{p}}=0.
\end{align} 
 Under the one parameter family of gauge transformations\footnote{In the following discussion, unless otherwise specified, by the phrases `gauge transformation' and `gauge-invariant', we mean, respectively, this one parameter family of gauge transformations and invariance under the same.} given by $A_a\rightarrow\tilde{A}_a=\{0,0,0,\tilde{A}_z(\eta)\}=\{0,0,0,A_z(\eta)+C\}$, where $C$ is a constant, the complex scalar field transforms to $\tilde{\Psi}=e^{iqCz}\Psi$.  This, in turn, implies that the Fourier transforms of $\tilde{\Psi}$ and $\Psi$ are related by $\tilde{\Psi}_{(p_z,\bf p_{\perp})}=\Psi_{(\tilde{p}_z,\bf p_{\perp})}$, where $\tilde{p}_z= p_z-qC$.  Therefore, the 3-vector $\bf p$ does not specify a given physical mode of the complex scalar field in a gauge-invariant way. On the other hand, the `physical momentum' 3-vector $\mathbf{\Pi}(\eta_0)\equiv(\Pi_{z}(\eta_0) = p_z+qA_z(\eta_0),\bf p_{\perp})$, where $\eta_0$ is an arbitrary time, can be used to specify a physical mode in a gauge-invariant manner, since, under a gauge transformation the quantity $p_{z}+qA_{z}(\eta_0)$ remains invariant. Once the gauge is fixed, however, we can always work with $\mathbf{p}$, with out any ambiguity.

It is evident that \ref{eomcomplex} describes a time dependent harmonic oscillator of unit mass and time-dependent frequency $\Omega_{\mathbf{p}}(\eta)$, where
\begin{align}\label{freqschwinger}
\Omega_{\mathbf{p}}^2(\eta)=	m^2+p_{\perp}^2+\left\{p_z+qA_z(\eta)\right\}^2
\end{align}
Thus, the Fourier modes of both (i) the conformally coupled, massive, scalar field in an expanding background and (ii) the complex scalar field in a homogeneous electric field satisfy the equation for a time-dependent harmonic oscillator with unit mass. Consequently, for appropriate choice of the parameters and functional forms for $a(\eta)$ and $A_{z}(\eta)$,  \ref{diffforfrw} and \ref{eomcomplex} can be made mathematically identical. That is, a purely algebraic correspondence between the equations of motion of the two systems exist when the following identification is made:
\begin{align}\label{correspondance}
k^2+M^2a^2(\eta)\rightleftarrows m^2+p_{\perp}^2+\left\{p_z+qA_z(\eta)\right\}^2
\end{align}
This, in turn, implies that the solutions in one case can be mapped to that in the other. We stress that this is just a useful \textit{algebraic} correspondence, at the level of differential equations governing the Fourier modes, using the principle that same equations have the same solutions. (We do not imply any physical equivalence between the two systems; for example, parameters which appear in the two cases do not share similar physical interpretation.)  In the following subsections we will investigate what this mapping translates to, in terms of the particle production, using some special cases.
\subsection{Sauter type electric field and its Friedmann analogue}

As a warm-up exercise for the above correspondence between a time dependent electric field and an expanding universe, we start by discussing the Sauter type electric field. This is a time dependent electric field and has the following form:
\begin{align}
\mathbf{E}=\left(0,0,E_0\textrm{sech}^2(\lambda \eta)\right)~.
\end{align}
The corresponding vector potential, called the Sauter-type potential, is given by,
\begin{align}\label{defineSauter}
\mathbf{A}=-\left(\frac{E_0}{\lambda}\right)\tanh\left(\,\lambda\,\eta\right)\hat{\mathbf{z}}
\end{align}
where, $\lambda$ is a constant having dimensions of inverse of time. The differential equation for the Fourier modes of a complex scalar field in this background electric field can be exactly solved and hence, the particle production rate can be explicitly calculated. The result is given by (see for example, \cite{sauterexact1,sauterexact2}):
\begin{align}
n_{\mathbf{k}}=\frac{\cosh^2\Big[\frac{\pi}{\lambda}\sqrt{\left(\frac{qE_0}{\lambda}\right)^2-1}\Big]-\sinh^2\Big[\frac{\pi}{2\lambda}\left(\tilde{\omega}_+-\tilde{\omega}_{-}\right)\Big]}{\sinh\Big(\frac{\pi\tilde{\omega}_+}{2\lambda}\Big)\sinh\Big(\frac{\pi\tilde{\omega}_-}{2\lambda}\Big)}
\end{align}
where, the frequencies $\tilde{\omega}_{\pm}$ are defined as,
\begin{align}\label{tilde_freq}
\tilde{\omega}_{\pm}=\sqrt{m^2+p_{\perp}^2+\left(p_z\mp\frac{qE_0}{\lambda}\right)^2}~.
\end{align}

We will next discuss how the above expression for particle number also arises for a conformally coupled massive scalar field in an expanding universe, for  a certain choice for the scale factor $a(\eta)$. To keep the discussion somewhat general, we will start with the following form of the scale factor \cite{toymodelexact},
\begin{align}\label{scalefactorforcorr}
a^2(\eta)=A+B\tanh(\lambda \eta)+C\tanh^2(\lambda\eta)~,
\end{align}
where $A$, $B$ and $C$ are dimensionless constants. The number of produced particles in the asymptotic limit, for the scale factor presented in \ref{scalefactorforcorr}, has been explicitly worked out in \cite{toymodelexact}, and is given by:
\begin{align}\label{particlenumber}
n_{\mathbf{k}}=\frac{\cosh\left(2\pi\frac{\omega_-}{\lambda}\right)+\cosh\left(\pi\sqrt{\frac{4M^2C}{\lambda^2}-1}\right)}{\cosh\left(2\pi\frac{\omega_+}{\lambda}\right)-\cosh\left(2\pi\frac{\omega_-}{\lambda}\right)}~,
\end{align}
where, the frequencies $\omega _{\pm}$ are defined in terms of the constants $A$, $B$ and $C$  appearing in the scale factor and the mass $M$ as,
\begin{align}
\omega_{\pm}=\frac{1}{2}\left\{\sqrt{k^2+M^2(A+B+C)}\pm\sqrt{k^2+M^2(A-B+C)}\right\}~.
\end{align}
To algebraically map the time dependent frequency $\omega_{\mathbf{k}}(\eta)$, with the scale factor $a(\eta)$ as in \ref {scalefactorforcorr}, to $\Omega_{\mathbf{p}}(\eta)$ for the Sauter-type potential, we start by making the following choice for the arbitrary constants, 
\begin{align}\label{choice}
A=\left(\frac{p_z}{m}\right)^2;\qquad B=-2\sqrt{AC};\qquad C=\left(\frac{qE_0}{m\lambda}\right)^2.
\end{align}
The scale factor, after this identification, becomes $a(\eta)=(p_{z}/m)-(qE_{0}/m)\tanh (\lambda \eta)$. (Note that in the expression for scale factor, $p_z$ now appears purely as a parameter that describes a family of scale factors.) Given this, one can immediately verify, following \ref{correspondance}, that with the identification $k^2=M^2+p_{\perp}^2$, the corresponding vector potential takes the form as in \ref{defineSauter}. Once this correspondence is established, we can use  \ref{particlenumber} and \ref{choice}, to obtain the particle number in an expanding universe:
\begin{align}
n_{\mathbf{k}}=\frac{\cosh\Big[\frac{\pi}{\lambda}\left(\tilde{\omega}_+-\tilde{\omega}_{-}\right)\Big]+\cosh\Big[\frac{2\pi}{\lambda}\sqrt{\left(\frac{qE_0}{\lambda}\right)^2-1}\Big]}{\cosh\Big[\frac{\pi}{\lambda}\left(\tilde{\omega}_++\tilde{\omega}_{-}\right)\Big]-\cosh\Big[\frac{\pi}{\lambda}\left(\tilde{\omega}_+-\tilde{\omega}_{-}\right)\Big]}
\end{align} 
This coincides exactly with the number of particles produced in a Sauter potential given in \ref{defineSauter} and the frequencies $\tilde{\omega}_{\pm}$ are those presented in \ref{tilde_freq}. This explicitly demonstrates how one may use the correspondence expressed by \ref{correspondance} to obtain the particle production rate in a given expanding background using the information about  particle production rate in a homogeneous electric field and vice versa. 

Before concluding this discussion, let us briefly discuss some of the limiting cases of the particle number in \ref{particlenumber}. First, consider the case $B=0=C$, in which case the scale factor is a constant. This yields, $\omega _{+}=\sqrt{k^{2}+M^{2}A}$ and $\omega _{-}=0$, so that the particle production rate vanishes, as it should:
\begin{align}\label{particlenumberlimit01}
\lim _{B,C \rightarrow 0}n_{\mathbf{k}}=\frac{1+\cosh\left(i\pi\right)}{\cosh(2\pi\frac{\omega_+}{\lambda})-1}=0.
\end{align}
Second, consider the vanishing $\lambda$ limit, when the  scale factor becomes $a(\eta)\sim A+ \lambda B\eta+\lambda ^{2}C\eta ^{2}$ and hence $a(\eta)$ becomes constant as $\lambda\rightarrow 0$. Thus, it is expected that the particle production should also vanish in the $\lambda \rightarrow 0$ limit. A key question is whether it vanishes  in an analytical fashion or non-analytically in $\lambda$. For positive $A$, $B$ and $C$, we obtain $\omega _{+}>\omega _{-}$ and in the vanishing $\lambda$ limit we have $\cosh(2\pi \omega _{\pm}/\lambda)\sim \exp(2\pi \omega _{\pm}/\lambda)$, as well as $\sqrt{(4M^{2}C/\lambda ^{2})-1}\sim (2M/\lambda)\sqrt{C}$. Thus, in the  $\lambda\to0$ limit, the particle number presented in \ref{particlenumber} becomes, 
\begin{align}\label{particlenumberlimit02}
n_{\mathbf{k}}
&\sim \exp\left\{-\frac{2\pi}{\lambda}\left(\omega_{+}-\omega _{-}\right)\right\}+\exp\left\{-\frac{2\pi}{\lambda}\left(\omega _{+}-M\sqrt{C} \right)\right\}
\end{align}
Since $\omega _{+}$ is greater than $\omega _{-}$ as well as $M\sqrt{C}$, it follows that $n_{\mathbf{k}}\rightarrow 0$ in the $\lambda \rightarrow 0$ limit. However, due to $(1/\lambda)$ dependence in the exponential, it is \textit{non-analytic} in $\lambda$ near $\lambda \sim 0$ as evident from \ref{particlenumberlimit02}. 

Finally, let us consider the other extreme, namely $\lambda\rightarrow\infty$ limit. The scale factor in this limit takes the following form
\begin{align}\label{signscalefactor}
a(\eta)=\sqrt{A+B\,\,\textrm{Sign}(\eta)+C}
\end{align}
where the `Sign' function is given by, $\textrm{Sign}(x)=1$ for $x>0$ and $\textrm{Sign}(x)=-1$ for $x<0$. (This scale factor corresponds to a `sudden' expansion at a single epoch.) Following an identical procedure, the particle number in this limit becomes
\begin{align}\label{pertcase}
n_{\mathbf{k}}=\frac{\omega_{-}^2}{\omega_{+}^2-\omega_{-}^2}+\frac{\left(\pi^2\omega_{+}^2\omega_{-}^2-3M^4C^2\right)}{3\lambda^2\left(\omega_{+}^2-\omega_{-}^2\right)}+\mathcal{O}(\lambda^{-4})
\end{align}
As evident from the above expression, the particle number $n_{\mathbf{k}}$ is \textit{analytic} near $\lambda\sim\infty$. Thus, we conclude that the particle production in an expanding universe for the scale factor in \ref{scalefactorforcorr} may be either analytic or non-analytic in $\lambda$ depending upon the limit of $\lambda$ under consideration. 

The behaviour of particle production rate $n_{\mathbf{k}}$ for the last two cases, namely $\lambda\rightarrow 0$ and $\lambda\rightarrow\infty$, is closely related to two well-known limiting cases of the Sauter potential. They are, respectively, the following two limits of \ref{defineSauter}: (i) $m\lambda/(qE_{0})\rightarrow 0$, i.e., when the electric field approaches a constant value --- the particle production rate in this case is non-analytic in $qE_0$; (ii) $m\lambda/(qE_{0})\rightarrow \infty$, i.e., when the electric field approaches a sharply localized function in time, like a pulse --- the particle production rate in this case is analytic in $qE_0$. Our mapping allows us to obtain an expanding universe analogue of these limits. 

In general, there is no assurance that the scale factor corresponding to an arbitrary electric field configuration is sourced by a physically acceptable matter distribution; this is the situation, for example, in the case for a localized pulse-like electric field obtained from $m\lambda/(qE_{0})\rightarrow \infty$ limit of the Sauter-type field. However, it turns out that the constant electric field --- with non-analytic dependence --- actually maps to a radiation dominated universe. We shall briefly discuss this situation next.
\subsection{Radiation dominated universe is equivalent to constant electric field}

As a second example, we examine whether a Friedmann universe with a scale factor, which can be generated by physically acceptable source, can be mapped to a constant electric field. For this purpose, let us again consider the $\lambda\rightarrow 0$ limit of scale factor in \ref{scalefactorforcorr}, but with the following choices of the parameters $B$ and $C$, such that
\begin{align}\label{parameter}
B=\frac{a_0\sqrt{A}}{\lambda};
\qquad
C=\frac{a_0^2}{4\lambda^2} 	
\end{align}
Here, $a_{0}$ is a constant with dimension of inverse length. In this particular case, after substitution of the previous expressions for $B$ and $C$, the square of the scale factor becomes 
\begin{align}\label{radmetric}
a^{2}(\eta)=\lim\limits_{\lambda\rightarrow 0}\left(A+\frac{a_0\sqrt{A}}{\lambda}\tanh (\lambda \eta)+\frac{a_0^2}{4\lambda^2}\tanh^2 (\lambda \eta)\right)=\left(\sqrt{A}+\frac{a_0\eta}{2}\right)^2.
\end{align} 
Therefore, the scalar factor in the conformal time coordinate has the form  $a(\eta)=\sqrt{A}+(a_0\eta/2)$. To see what kind of matter fluid may generate the same, let us consider the scale factor to be sourced by an ideal fluid with the equation of state $p=w\rho$, where $p$ is the pressure and $\rho$ is the energy density, then the scale factor evolves with the conformal time as:
\begin{align}
a(\eta)=(b_{0}+b_{1}\eta)^{\frac{2}{(1+3w)}}
\end{align} 
where $b_{0}$ and $b_{1}$ are two unknown constants of integration.  For a radiation dominated universe, we have $w=1/3$, so that the scale factor becomes $a(\eta)\propto \eta$, which immediately connects to the scale factor given by \ref{radmetric}. Thus the choices made in \ref{parameter} for  the  constants $B$ and $C$  corresponds to a radiation dominated universe. In this case, we have the following limiting behaviour for the particle number in the asymptotic limit, (see \ref{radiation} for details),
\begin{align}\label{radparticle}
\lim _{\lambda \rightarrow 0}n_{\mathbf{k}}&
=\exp\left(-\frac{2\pi k^2}{Ma_0}\right)
\end{align}
where $k^{2}=|\mathbf{k}|^{2}$. When $a_0M\rightarrow 0$, we expect the particle production to vanish because: (i) as $a_0\rightarrow 0$, the spacetime is flat (ii) as $M\rightarrow 0$, because of the conformal coupling, the scalar field in the background of a flat Friedmann metric is equivalent to that in Minkowski spacetime. The particle production rate indeed drops to zero as $a_0\rightarrow 0$ in \ref{radparticle}, but in a \textit{non-analytic} fashion. 
Identical scenario arises in the context of constant electric field as well, where the particle number is  non-analytic in the coupling constant. This analogy can in fact be made more precise in the light of correspondence given in \ref{correspondance}, with the following identification:
\begin{align}
k^2\rightleftarrows p_{\perp}^2+m^2;\qquad M\sqrt{A}\rightleftarrows p_z;\qquad \frac{M a_0}{2} \rightleftarrows qE_0
\end{align}
so that the particle number  in \ref{radparticle} takes the following familiar form
\begin{align}
n_{\mathbf{p}}=\exp\left[-\frac{\pi(p_{\perp}^2+m^2)}{qE_0}\right]
\end{align}
The particle number $n_{\mathbf{p}}$, clearly, matches that in the context of a constant electric field and is non-analytic in $qE_0$. It is interesting to see that two of the important cases of pair production, namely the Schwinger effect and pair creation in a radiation dominated universe, which are seemingly different, are related in a very simple manner \cite{Padmanabhan:1991uk,toymodelexact,Rajeev2018}. Next, we will seek for a time dependent electric field configuration that corresponds to a de Sitter or quasi-de Sitter spacetime.
\subsection{de Sitter universe is equivalent to a singular electric field}\label{singularfield}

The above analysis shows that the well-known case of Schwinger effect can be mapped to a radiation dominated universe and the non-analytic behaviour of the particle number holds true in the radiation dominated universe as well. We will next discuss the mapping in the reverse direction i.e., we will start from a well-known expanding universe, namely de Sitter, and then study the form of the   electric field it maps to. To keep the discussion slightly general, will start with a generalization of the de Sitter spacetime, described by the following scale factor:
\begin{align}\label{pseudods}
a(\eta)=\left(a_0+\frac{1}{1-H\eta}\right)
\end{align}
This metric approaches: (i) the Minkowski metric, except for some rescaling, when $|H\eta|\gg 1$ and (ii) the de Sitter metric as $H\eta\approx 1$ or as $a_{0}\rightarrow 0$. From the Friedmann equations, we can determine the density $\rho$ and pressure $p$ of the ideal fluid that can act as the source for this geometry. They  are given by:
\begin{align}
\rho(a)&\equiv\frac{3}{8\pi G}\frac{1}{a^{4}}\left(\frac{da}{d\eta}\right)^{2}=\frac{3H^2}{8\pi G}\frac{(a-a_{0})^4}{a^4};
\nonumber
\\
p(a)&\equiv -\frac{1}{8\pi G}\left\{3\frac{(da/d\eta)^{2}}{a^{4}}+\frac{2}{a}\frac{d}{d\eta}\left(\frac{(da/d\eta)}{a^{2}}\right) \right\} =-\frac{3H^2}{8\pi G}\frac{(a+a_0/3)(a-a_0)^3}{a^4}~.
\end{align}
The density and pressure vanishes as $a\rightarrow a_0$ (equivalently, as $\eta\rightarrow-\infty$), as expected, since the spacetime approaches Minkowski space in this limit. On the other hand, as $a\rightarrow\infty$ (equivalently, as $\eta\rightarrow H^{-1}$), the density and pressure approaches constant values such that $\rho=-p=(3H^2)/(8\pi G)$, like in the de Sitter spacetime.

Now, from \ref{diffforfrw}, we find that the Fourier modes of a massive conformally coupled scalar field in this background satisfies the following differential equation:
\begin{align}\label{difffords}
\Phi_{\mathbf{k}}''+\left[k^2+M^2\left(a_0+\frac{1}{1-H\eta}\right)^2\right]\Phi_{\mathbf{k}}=0~.
\end{align}
We can simplify this differential equation by introducing two new parameters $\kappa$ and $\mu$, as well as a new variable $z$, such that,
\begin{align}\label{parameters}
\kappa=\frac{i a_0 M^2}{H \sqrt{a_0^2 M^2+k^2}};\qquad
\mu=\sqrt{\frac{1}{4}-\frac{M^2}{H^2}};\qquad
z=\frac{2 i (H t-1) \sqrt{a_0^2 M^2+k^2}}{H}~.
\end{align}
In terms of these variables, \ref{difffords} takes the form,
\begin{align}\label{diffinWhitform1}
\frac{d^2\Phi_{\mathbf{k}}}{dz^2}+\left(-\frac{1}{4}+\frac{\kappa}{z}+\frac{\frac{1}{4}-\mu^2}{z^2}\right)\Phi_{\mathbf{k}}=0~,
\end{align}
The solutions to this differential equation can be written in terms of Whittaker functions $W_{\kappa,\mu}(z)$ and $M_{\kappa,\mu}(z)$. In particular, the `in'-modes, which are the solutions to \ref{diffinWhitform1} that behave as positive frequency functions near $\eta\rightarrow-\infty$, are given by $W_{\kappa,\mu}(z)$. One can verify this by looking at the behaviour of $\phi_{\mathbf{k}(\rm{in})}$ near the asymptotic past:
\begin{align}
\phi_{\mathbf{k}({\rm in})}\sim e^{-i(a_0^2 M^2+k^2)^{1/2}\eta};\qquad
\eta\rightarrow-\infty
\end{align}
The `out'-modes, on the other hand, are oscillatory and thus well-defined only when $(M^2/H^2)>1/4$, a result which arises repeatedly in the context of de Sitter spacetime. (We will comment on this feature, in detail, later on.) The parameter $\mu$ becomes purely imaginary in this case, so that we can write $\mu=i|\mu|$ and the `out'-modes $\phi_{\mathbf{k}({\rm out})}$ turns out to be proportional to $M_{\kappa,i|\mu|}(z)$. From the asymptotic expansion of the Whittaker function, it follows that, the `out' modes take the following form at late times:
\begin{align}
\phi_{\mathbf{k}({\rm out})}\sim e^{-i|\mu|H t};\qquad \eta\rightarrow H^{-1}
\end{align}
where, $t$ is the cosmic time and related to the conformal time $\eta$ through the well known relation, $dt=a(\eta)d\eta$. The `in'-modes and `out'-modes introduced above are related through a Bogoliubov transformation of the following form:
\begin{align}
\phi_{\mathbf{k}({\rm out})}=\alpha_{\mathbf{k}} \phi_{\mathbf{k}({\rm in})}+ \beta_{\mathbf{k}} \phi_{\mathbf{k}({\rm in})}^{*}
\end{align}
where, $\alpha_{\mathbf{k}}$ and $\beta_{\mathbf{k}}$ are the Bogoliubov coefficients. To find the explicit expressions for $\alpha_{\mathbf{k}}$ and $\beta_{\mathbf{k}}$, we can use the following relation involving the Whittaker functions,
\begin{align}
M_{\kappa,\mu}(z)&=\frac{\Gamma(2\mu+1)e^{i\pi(\kappa-\mu-\frac{1}{2})}}{\Gamma\left(\mu+\kappa+\frac{1}{2}\right)}W_{\kappa,\mu}(z)+\frac{\Gamma(2\mu+1)e^{i\pi\kappa}}{\Gamma\left(\mu-\kappa+\frac{1}{2}\right)}W_{-\kappa,\mu}(-z)
\\
&\equiv \mathcal{A}_{\mathbf{k}}\,\,W_{\kappa,\mu}(z)+\mathcal{B}_{\mathbf{k}}\,\,W_{-\kappa,\mu}(-z)
\end{align}
where, the last line defines the constants $\mathcal{A}_{\mathbf{k}}$ and $\mathcal{B}_{\mathbf{k}}$ respectively. It is then straightforward to see that the Bogoliubov coefficients are given in terms of the constants $\mathcal{A}_{\mathbf{k}}$ and $\mathcal{B}_{\mathbf{k}}$ such that,
\begin{align}
\alpha_{\mathbf{k}}=\frac{\mathcal{A}_{\mathbf{k}}}{\sqrt{|\mathcal{A}_{\mathbf{k}}|^2-|\mathcal{B}_{\mathbf{k}}|^2}};&&\beta_{\mathbf{k}}=\frac{\mathcal{B}_{\mathbf{k}}}{\sqrt{|\mathcal{A}_{\mathbf{k}}|^2-|\mathcal{B}_{\mathbf{k}}|^2}}.
\end{align}
Further, the number (density) of particles produced in the asymptotic future can be expressed as
\begin{align}\label{nforpseudods}
n_{\mathbf{k}}=|\beta_{\mathbf{k}}|^2=e^{-\pi(|\kappa|+|\mu|)}\frac{\cosh\left\{\pi(|\kappa|-|\mu|)\right\}}{\sinh\left\{2\pi|\mu|\right\}}
\end{align}
Recall that, the $a_0\rightarrow 0$ limit of the scale factor in \ref{pseudods} describes an exact de Sitter spacetime. The number of particles produced in this limit is given by
\begin{align}
\lim_{a_0\rightarrow 0}n_{\mathbf{k}}=\frac{1}{e^{2\pi|\mu|}-1}
\end{align}
This matches with the well known result in literature \cite{schwingerinds1,schwingerinds3} and we will explore a closely related case further in \ref{dsbogolsec}. In the limit $H\rightarrow 0$, the scale factor becomes constant and hence we expect zero particle production. This comes out naturally from \ref{nforpseudods}, since the RHS in the $|\mu|\rightarrow \infty$ (equivalent to $H\rightarrow 0$ limit) identically vanishes. We also see that this vanishing occurs through a non-analytic dependence in the parameter $H$. On the other hand, the above formula is not applicable for $M\rightarrow 0$ limit, as the  mode functions at late times will not be of oscillatory nature in this case. (We will discuss this feature in detail later on.) One can, of course, work out the $M=0$ case separately and prove that the particle production vanishes. This is consistent, since there cannot be any particle production for massless,  conformally coupled scalar field in a conformally flat spacetime.

We will now determine an electric field configuration that, in accordance with \ref{correspondance}, corresponds to the quasi-de Sitter metric that we have introduced. The form of scale factor  suggests that the vector potential will be of the following form:
\begin{align}
A_{z}(\eta)=\frac{E_0}{\omega(1-\omega \eta)}-\frac{E_0}{\omega}
\end{align}
By demanding that the time dependent frequencies $\omega_{\mathbf{k}}(\eta)$ and $\Omega_{\mathbf{p}}(\eta)$ to be algebraically same, we arrive at the following identification between the parameters of the two scenarios,
\begin{align}
p_z-\frac{qE_0}{\omega}=Ma_0;\qquad m^2+p_{\perp}^2=k^2;\qquad \omega=H;\qquad \frac{qE_0}{\omega}=M~.
\end{align}
The time dependent electric field turns out to be singular at $\eta=\omega^{-1}$, where it diverges quadratically. Its explicit form is given by
\begin{align}
\mathbf{E}=\left(0,0,\frac{E_0}{(1-\omega\eta)^2}\right)
\end{align}
This describes a family of electric fields parametrized by $E_0$ and $\omega$. (It is tempting to interpret the parameter $\omega$ as the inverse of time at which the electric field diverges. But, by virtue of a shift in the time coordinate, the point of divergence can be shifted to any arbitrary instant in time.) It turns out that the above electric field satisfies the following condition,
\begin{align}
\frac{1}{4}\frac{(\partial_{\eta}|\mathbf{E}|)^2}{|\mathbf{E}|^3}=\textrm{constant}=\frac{\omega^{2}}{E_{0}}\equiv \sigma
\end{align} 
Let us now determine the  number of particles $n_{\mathbf{p}}$, associated with a complex scalar field, produced in this electric field background during the period $-\infty<\eta <\omega^{-1}$.  First, note that the parameters $\kappa$ and $\mu$ defined in \ref{parameters}, are replaced by the following expressions,
\begin{align}\label{dS_Corres}
\kappa=\frac{iqE_0 \left(p_z-qE_0\omega^{-1}\right)}{\omega^2\sqrt{(p_z-qE_0\omega^{-1})^2+p_{\perp}^2+m^2}};\qquad
\mu=\sqrt{\frac{1}{4}-\frac{q^{2}}{\sigma^{2}}}
\end{align}
and the condition, $\mu=i|\mu|$ translates to $\sigma<2q$. Having identified the parameters that are related to each other in either side of the correspondence, we can use \ref{nforpseudods} to show that the particle number for the above time dependent electric field becomes,
\begin{align}\label{nforsingular}
n_{\mathbf{p}}&=\frac{\cosh\left[\pi\left(\frac{qE_0 \left(p_z-qE_0\omega^{-1}\right)}{\omega^2\sqrt{(p_z-qE_0\omega^{-1})^2+p_{\perp}^2+m^2}}-\sqrt{\frac{(qE_0)^2}{\omega^4}-\frac{1}{4}}\right)\right]}{\sinh\left[2\pi\sqrt{\frac{(qE_0)^2}{\omega^4}-\frac{1}{4}}\right]}
\nonumber
\\
& \qquad \qquad  \times  \exp\left[-\pi\left(\frac{qE_0 \left(p_z-qE_0\omega^{-1}\right)}{\omega^2\sqrt{(p_z-qE_0\omega^{-1})^2+p_{\perp}^2+m^2}}+\sqrt{\frac{(qE_0)^2}{\omega^4}-\frac{1}{4}}\right)\right]
\end{align}
The following aspects are worth noticing as regards this result: 

(a) In the $\omega\rightarrow 0$ limit, electric field approaches a constant and the particle number density $n_{\mathbf{k}}$ approaches the Schwinger's result, which is expected. 

(b) The $p_z\rightarrow qE_0/\omega$ limit (which is  the analogue of pure de Sitter spacetime) gives
\begin{align}\label{part_e_omega}
\lim_{p_z\rightarrow (qE_{0}/\omega)}n_{\mathbf{p}}=\frac{1}{\exp\left[2\pi\sqrt{\frac{(qE_0)^2}{\omega^4}-\frac{1}{4}}\right]-1}
\end{align}
Let us discuss the equivalent of $H\rightarrow 0$ limit in this context. As the correspondence in \ref{dS_Corres} shows, this is achieved by taking the following two limits: $\omega \rightarrow 0$ as well as $qE_{0}\rightarrow 0$, keeping $M=(qE_{0}/\omega)$ finite. As evident from \ref{part_e_omega}, the particle number identically vanishes in this limit, which is what we expect in the limit of vanishing electric field. 

(c) On the other hand, the above estimation for particle number density is not applicable for $qE_{0}\rightarrow 0$ limit, as $\sigma$ diverges, rendering the above analysis inapplicable. This is identical to the massless, conformally coupled,  limit of de Sitter. The particle number indeed vanishes in this limit, but this case needs to be worked out separately.  

(d) It is instructive to rewrite \ref{nforsingular} in a gauge-invariant manner. In order to do that, we first define the gauge-invariant `physical momentum' at the asymptotic past by $\mathbf{\Pi}=(p_z+qA_z(-\infty),\bf p_{\perp})$. It is easy to see that $\Pi_z=(p_z-qE_0/\omega)$ and $\mathbf{\Pi}_{\perp}=\mathbf{p}_{\perp}$. This motivates us to define an `energy' for each mode by $\epsilon_{\mathbf{p}}=\sqrt{\Pi^2+m^2}$, where $\Pi^2=|\mathbf{\Pi}|^2$. The particle production rate can then be written as
\begin{align}\label{nforsingular2}
n_{\mathbf{p}}&=\frac{\cosh\left(\frac{\pi q\Pi_{z}}{\sigma\epsilon_{\mathbf{p}}}-\pi|\mu|\right)}{\sinh\left(2\pi|\mu|\right)}
\exp\left[-\frac{\pi q\Pi_{z}}{\sigma\epsilon_{\mathbf{p}}}-\pi|\mu|\right]
\end{align}
When $\Pi\gg m$, i.e., in the ultra-relativistic limit, the above expression approximates to
\begin{align}\label{nforsingular3}
n_{\mathbf{p}}&\approx\frac{\cosh\left(\frac{\pi q\cos\theta}{\sigma}-\pi|\mu|\right)}{\sinh\left(2\pi|\mu|\right)}
e^{-\left(\frac{\pi q\cos\theta}{\sigma}+\pi|\mu|\right)};\qquad\Pi\gg m,
\end{align}
where, $\theta$ is the angle between electric and the `physical momentum' $\mathbf{\Pi}$. It is interesting to note that, in the ultra-relativistic limit, the leading order particle production depends only on the direction of `physical momentum' and is independent of its magnitude.

To summarize, we have shown that the particle production by a scalar field in an expanding Friedmann spacetime, that smoothly extrapolates from Minkoswki space to de Sitter universe, can be algebraically mapped to that of a complex scalar field in the background of a singular electric field that diverges quadratically at a certain instant of time. The special case of particle production in de Sitter spacetime (i.e., $a_0=0$), under this map, translates to the case of particle production by the complex scalar field with a certain value of the component of canonical momentum along the electric field (i.e., $p_{z}=qE_0\omega^{-1}$). 
\subsection{An electric field that produces Planck spectrum of particles} 

The background geometries which produce a Planck spectrum of particles (e.g., black hole spacetimes) are of considerable importance in the study of quantum field theory in curved spacetime. This prompts us to ask: Is there a time dependent electric field in the flat spacetime which produces a Planck spectrum of particles? 

It is well-known that, for a suitable vacuum choice, the Milne universe does lead to a Planck spectrum of particles at late times \cite{toymodelexact,PhysRevLett.64.2471}. Therefore, the corresponding electric field will lead to the same result. We briefly mention this result here, postponing detailed discussion of this `black hole analogue model' to a future work \cite{scrktp}. The scale factor for Milne universe is given by $a(t)=\mathcal{H}t$ where $\mathcal{H}$ has inverse dimensions of time. 
(The standard Hubble parameter is  $(\dot a/a)=1/t$ and hence $\mathcal{H}$ is \textit{not} the Hubble parameter.)
The passage to conformal time is straightforward and one obtains, $\mathcal{H}t=\exp(\mathcal{H}\eta)$, such that $t=1/\mathcal{H}$ corresponds to $\eta=0$ and $t=0$ relates to $\eta=-\infty$. Thus the scale factor in conformal time reads, $a(\eta)=\exp(\mathcal{H}\eta)$. 
As we will see it is convenient to generalise the discussion slightly and consider the scale factor  of the following form:
\begin{align}\label{milnelike}
	a(\eta)=a_0+e^{\mathcal{H}\eta}.
\end{align}
Notice that the scale factor reduces to that of Milne universe in the $a_0\rightarrow 0$ limit. Moreover, the spacetime corresponding to the above scale factor smoothly extrapolates from a Minkowski space time (when, $\eta<0$ and $|\mathcal{H}\eta|\gg 1$) to a Milne universe (when, $\eta>0$ and $|\mathcal{H}\eta|\gg 1$). For a pure Milne universe (i.e., $a_0=0$), it can be shown that the density of produced particles \cite{toymodelexact,PhysRevLett.64.2471} is Planckian: 
\begin{align}\label{milne_1}
n_{\mathbf{k}}=\frac{1}{\exp\left(\frac{2\pi k}{\mathcal{H}}\right)-1}
\end{align}
with the temperature given by $T=\mathcal{H}/(2\pi)$.

To study the particle production in the `generalized Milne' universe, corresponding to the scale factor in \ref{milnelike}, let us first consider the equation of motion of the Fourier mode $\Phi_{\bf k}$ in this background, which is given by
\begin{align}\label{diffinmilne}
	\Phi_{\bf k}''+\left[k^2+M^2(a_0+e^{\mathcal{H}\eta})^2\right]\Phi_{\bf k}=0.
\end{align}
It is convenient at this stage to define a new dependent variable $\xi_{\mathbf{k}}$, such that
\begin{align}
	\Phi_{\bf k}=e^{-\frac{1}{2}\mathcal{H}\eta}\xi_{\mathbf{k}}
\end{align}
and a new independent variable $z$ by
\begin{align}
	z=\frac{2ie^{\mathcal{H}\eta}M}{\mathcal{H}},
\end{align}
so that, \ref{diffinmilne} reduces the standard form of Whittaker's differential equation:
\begin{align}
	\frac{d^2\xi_{\mathbf{k}}}{dz^2}+\left(-\frac{1}{4}+\frac{1/4-\mu^2}{z^2}\right)\xi_{\mathbf{k}}=0.
\end{align}
Hence, the general solution to \ref{diffinmilne} can be written in terms of the Whittaker functions $W_{\kappa,\mu}(z)$ and $M_{\kappa,\mu}(z)$. It turns out that, the out modes $\phi_{\mathbf{k}(\rm out)}$ is given by
\begin{align}
	\phi_{\mathbf{k}({\rm out})}\propto e^{-\frac{1}{2}\mathcal{H}\eta}W_{\kappa,\mu}(z).
\end{align}
One can verify this by noting that the late time behaviour of $\phi_{\mathbf{k}({\rm out})}$ turns out to be:
\begin{align}
	\phi_{\mathbf{k}({\rm out})}\propto e^{-\frac{1}{2}\mathcal{H}\eta}e^{-i\frac{M}{\mathcal{H}}e^{\mathcal{H}\eta}+i\frac{a_0M}{\mathcal{H}}\eta};\qquad\eta\rightarrow\infty.
\end{align} 
Clearly, this mode behaves as a positive frequency solution at late times and hence, qualify as the `out-modes'. On the other hand, in the early times, $\phi_{\mathbf{k}({\rm out})}$ has the following limiting behaviour:
\begin{align}\label{earlymilnemode}
	\phi_{\mathbf{k}({\rm out})}\propto \tilde{\alpha}_{\mathbf{k}}e^{-i\sqrt{k^2+M^2a_0^2}\eta}+\tilde{\beta}_{\mathbf{k}}e^{i\sqrt{k^2+M^2a_0^2}\eta};\qquad\eta\rightarrow-\infty,
\end{align}
where,
\begin{align}
	\tilde{\alpha}_{\mathbf{k}}=e^{\frac{\pi|\mu|}{2}}\frac{\Gamma\left(2\mu\right)}{\Gamma\left(\mu-\kappa+\frac{1}{2}\right)};&&\tilde{\beta}_{\mathbf{k}}=e^{-\frac{\pi|\mu|}{2}}\frac{\Gamma\left(-2\mu\right)}{\Gamma\left(-\mu-\kappa+\frac{1}{2}\right)}.
\end{align}
The asymptotic value of particle production rate can then be evaluated to get:
\begin{align}\label{partilcespectrum}
	n_{\mathbf{k}}=\frac{|\tilde{\beta}_{\mathbf{k}}|^2}{|\tilde{\alpha}_{\mathbf{k}}|^2-|\tilde{\beta}_{\mathbf{k}}|^2}=\frac{e^{\frac{2\pi}{\mathcal{H}}\left(a_0M+\sqrt{k^2+M^2a_0^2}\right)}+1}{e^{\frac{4\pi}{\mathcal{H}}\sqrt{k^2+M^2a_0^2}}-1}
\end{align}
From \ref{earlymilnemode}, we can see that the `in-mode' labelled by $\mathbf{k}$ has the energy $\epsilon_{\mathbf{k}}=\sqrt{k^2+M^2a_0^2}$. Hence, in the ultra-relativistic limit, given by $\epsilon_{\mathbf{k}}\gg Ma_0$, the particle spectrum in \ref{partilcespectrum} approximates to
\begin{align}
	n_{\mathbf{k}}\approx\frac{1}{e^{\frac{2\pi}{\mathcal{H}}\epsilon_{\mathbf{k}}}-1}.
\end{align}
This corresponds to a Planckian distribution with the temperature $T=\mathcal{H}/(2\pi)$. Note that, in the special case of $a_0=0$, which corresponds to the pure Milne universe, the spectrum is \textit{exactly} Planckian. 

The correspondence of scale factor in \ref{milnelike} with an electric field can be easily achieved, by virtue of \ref{correspondance}, which gives,
\begin{align}\label{milne_correspondence}
k^{2}=m^{2}+p_{\perp}^{2};\qquad qA_{z}(\eta)+p_{z}=M(a_0+e^{\mathcal{H}\eta})
\end{align}
Hence, the associated electric field becomes,
\begin{align}\label{milne_efield}
E_{z}(\eta)=E_{1}\exp(\mathcal{H}\eta);\qquad E_{1}=\frac{M\mathcal{H}}{q}.
\end{align}
Thus, at $\eta \rightarrow -\infty$, we obtain $E(\eta)=0$, while for $\eta=0$, we have $E(\eta)=E_{1}$. Therefore, using \ref{partilcespectrum} and \ref{milne_correspondence}, the number density of the quanta of a complex scalar field, produced due to the coupling with this time dependent electric field is given by:
\begin{align}\label{milne_2}
n_{\mathbf{p}}=\frac{e^{\frac{2\pi}{\mathcal{H}}\left(\Pi_z+\epsilon_{\mathbf{p}}\right)}+1}{e^{\frac{4\pi}{\mathcal{H}}\epsilon_{\mathbf{p}}}-1},
\end{align}
where, we have defined the gauge-invariant `physical momentum' $\mathbf{\Pi}$ and the energy $\epsilon_{\mathbf{p}}$ for each modes, respectively, by $\mathbf{\Pi}\equiv(p_z+qA_z(-\infty),\mathbf{p}_{\perp})$ and $\epsilon_{\mathbf{p}}=\sqrt{\Pi^2+m^2}$. For small values of longitudinal physical momentum, i.e., for $\Pi_z\ll\epsilon_{\mathbf{p}}$, the particle spectrum approximates to
\begin{align}\label{thermalelectric}
	n_{\mathbf{p}}\approx\frac{1}{e^{\frac{2\pi}{\mathcal{H}}\epsilon_{\mathbf{p}}}-1},
\end{align}
which corresponds to a Planckian distribution with temperature $T=\mathcal{H}/(2\pi)$. The expression for $T$ is reminiscent of that of a fictitious de Sitter space time with Hubble parameter $\mathcal{H}$. (We also note that $T$ is  the Davies-Unruh temperature corresponding to the asymptotic acceleration $g=(qE_1/M)$. However,  $M$ is \textit{not} the mass of the complex scalar field under consideration, but that of the scalar field in the generalized Milne universe; so this  interpretation of $T$ as a Davies-Unruh temperature is incorrect.) The fact that a thermal spectrum can be generated from a homogeneous but time dependent electric field is an interesting results by itself and definitely needs further study \cite{scrktp}.

In the $\mathcal{H}\rightarrow 0$ limit, the electric field in \ref{milne_efield} approaches a constant. Therefore, we expect that the particle rate in the $\mathcal{H}\rightarrow 0$ limit, approaches the Schwinger's result. In order to see that this is indeed the case, we choose the vector potential to be of the following form:
\begin{align}\label{vectformilne}
	A_{z}(\eta)=-\frac{E_1}{\mathcal{H}}(e^{\mathcal{H}\eta}-1).
\end{align}
Clearly, the $\mathcal{H}\rightarrow0$ limit of this potential is given by $-E_1\eta$, as is desired. The explicit expression for the longitudinal `physical momentum' of a mode labelled by $\mathbf{p}$, in this gauge, becomes $\Pi_z=p_z+qE_1/\mathcal{H}$. The particle number density, given by \ref{milne_2}, can then be rewritten as
\begin{align}\label{milne_3}
n_{\mathbf{p}}=\frac{\exp\left\{\frac{2\pi}{\mathcal{H}}\left[(p_z+qE_1/\mathcal{H})^2+\sqrt{m^2+p_{\perp}^2+(p_z+qE_1/\mathcal{H})^2}\right]\right\}+1}{\exp\left[\frac{4\pi}{\mathcal{H}}\sqrt{m^2+p_{\perp}^2+(p_z+qE_1/\mathcal{H})^2}\right]-1}.
\end{align}
In the small $\mathcal{H}$ limit, the above expression, to the leading order, reduces to $n_{\mathbf{p}}=\exp{[-\pi(m^2+p_{\perp}^2)/(qE_1)]}$, which matches exactly with the Schwinger's result.

We will conclude this section with a brief comment regarding the correspondence between the electric field and more general expansion factors of the universe, for the sake of completeness. Consider a universe sourced by matter with the equation of state $p=w \rho$, with constant $w$ ($\neq -1$). Then the scale factor behaves as, $a(t)=(t/t_{0})^{\frac{2}{3(1+w)}}$. The conformal time is: 
\begin{align}\label{EoS_1}
\eta=\frac{3(1+w)}{1+3w}t_{0}^{\frac{2}{3(1+w)}}t^{\frac{1+3w}{3(1+w)}}
\end{align}
so that $t/t_{0}=(\eta/\eta_{0})^{3(1+w)/(1+3w)}$. The scale factor, in terms of the conformal time, is then $a(\eta)=(\eta/\eta_{0})^{2/(1+3w)}$. The mode functions then satisfy the following differential equation,
\begin{align}\label{tpEoS_2}
\Phi_{\mathbf{k}}''+\left\{\mathbf{k}^{2}+M^{2}\left(\frac{\eta}{\eta_{0}}\right)^{\frac{4}{(1+3w)}} \right\}\Phi_{\bf k}=0
\end{align}
The corresponding vector potential for the equivalent electric field is easy to find using
\ref{correspondance}. We get: 
\begin{align}
\mathbf{k}^{2}=m^{2}+p_{\perp}^{2};\qquad M\left(\frac{\eta}{\eta_{0}}\right)^{\frac{2}{(1+3w)}}=qA_{z}(\eta)+p_{z}
\end{align}
so that the electric field becomes,
\begin{align}
E_{z}=-\frac{2M}{(1+3w)q}\frac{1}{\eta_{0}}\left(\frac{\eta}{\eta_{0}}\right)^{\frac{1-3w}{1+3w}}
\end{align}
Thus most of our discussion can be generalized to this case as well, when the mode functions are known. Unfortunately, the closed form solution to \ref{tpEoS_2} is known only in a few special cases. Thus we will not pursue this analogy any further. We will next take up more general features suggested by the mapping between time-dependent electric field and the expanding universe.
\section{Perturbative vs non-perturbative limits of particle production}\label{section2}

In an earlier work \cite{Rajeev2018}, we studied pair production in a homogeneous electric field background with the emphasis on analytic vs non-analytic dependence of the asymptotic particle number on the coupling constant $q$. In that case, we could obtain two distinct general classes of electric field configurations that exhibit, respectively, analytic and non-analytic behaviour in the coupling constant. In this section, we will explore the implications of these results for particle production in an expanding universe using  the correspondence discussed in \ref{sec1}.

Recall that the time dependent harmonic oscillator equation satisfied by $\Phi_{\bf k}$ is given by  
\begin{align}\label{tdhofrw}
\partial^2_{\eta}\Phi_{\mathbf{k}}+k^2\left[1+\frac{a^2(\eta)}{\gamma^2}\right]\Phi_{\mathbf{k}}=0;\qquad \gamma=\frac{k}{M}
\end{align}
Let us now study the solutions of this equation in two regimes: (i) when $a^2/\gamma^2$ is ``small'' and can be considered as a perturbation and (ii) when $a^2/\gamma^2$ cannot be treated as a perturbation. The precise meaning of these conditions will become clear as we proceed.
\subsection{Perturbative limit}\label{pert}

Let us consider a regime of expansion when that the scale factor is bounded from above by some value, such that:
\begin{align}\label{condone}
a(\eta)\leq a_{\max}\ll \gamma
\end{align}
In this case, the time dependent term in \ref{tdhofrw} can be treated as a perturbation. For a Friedman universe expanding monotonically  from a singularity ($a=0$), there always exists an epoch in which this condition holds. The particle number  obtained may then be interpreted as the instantaneous particle number at the end this epoch. We can then  expand $\Phi_{\bf k}$ as
\begin{align}
	\Phi_{\bf k}(\eta)=\Phi_{\mathbf{k}(0)}+\frac{1}{\gamma}\Phi_{\mathbf{k}(1)}+\frac{1}{\gamma^2}\Phi_{\mathbf{k}(2)}+...
\end{align}
We seek  a solution $\phi_{\mathbf{k}}$ to \ref{tdhofrw} that behaves as $e^{-ik\eta}$ as $\eta\rightarrow -\infty$. Using standard perturbative analysis techniques we find that:
\begin{align}
	\Phi_{\bf k}(\eta)=e^{-ik\eta}-\frac{k}{\gamma^2}\int_{-\infty}^{\eta}d\eta'\sin\left[k(\eta-\eta')\right]a^2(\eta')e^{-ik\eta'}+\mathcal{O}(\gamma^{-4})
\end{align}
The asymptotic behaviour of this solution to leading order in $\gamma^{-1}$ is then given by
\begin{align}
	\phi_{\mathbf{k}}(\eta)=\begin{cases}
	e^{-ik\eta};\qquad \eta\rightarrow-\infty\\
	\mathcal{A}e^{-ik\eta}+\mathcal{B}e^{ik\eta};\qquad \eta\rightarrow\infty
	\end{cases}
\end{align}
where, $\mathcal{A}=1+\mathcal{O}(\gamma^{-2})$ and $\mathcal{B}=(i\pi k/\gamma^2)\chi (2k)$ with,
\begin{align}
	\chi(\mu)=\int_{-\infty}^{\infty}\frac{d\eta}{2\pi}a^{2}(\eta)e^{-i\mu\eta}
	\label{ft1}
\end{align}
being the Fourier transform of the conformal factor. The number of particles $n_{\mathbf{k}}$ produced at the asymptotic future, to leading order in $\gamma^{-1}$, can then be calculated as
\begin{align}\label{pertparticle}
	n_{\mathbf{k}}=|\mathcal{B}|^2=\left(\frac{\pi M^2}{k}\right)^2|\chi(2k)|^2+\mathcal{O}(\gamma^{-4}).
\end{align}

As an example, let us apply this result to the large $\lambda$ limit of \ref{scalefactorforcorr}, in which case the scale factor is given by \ref{signscalefactor}. In this case, the Fourier transform in \ref{ft1} can be easily evaluated to get $\chi(2k)=iB/(2k\pi)$. Hence the leading order particle number, from \ref{pertparticle} takes the form
\begin{align}
	n_{\mathbf{k}}=\left(\frac{BM^2}{2k^2}\right)^2+\mathcal{O}(\gamma^{-4})
\end{align}
One can easily verify that this is consistent with leading order behaviour of $n_{\mathbf{k}}$ given in \ref{pertcase}. 
\subsection{Non-perturbative limit}\label{nonpert}

We will next consider the more interesting case of the non-perturbative limit. The idea is to translate the procedure adopted in Ref.\cite{Rajeev2018}, for a time-dependent electric field,  to the expanding universe case. 
 This arises, when the scale factor is such that, $a(\eta)\gg \gamma$, as well as $|\eta|>\eta_c$. (That is, at some critical value of time $\eta=\eta_c$, the perturbative analysis, discussed in \ref{pert}, fails.) We will further assume that the scale factor is changing adiabatically in the asymptotic past and future, i.e.,
\begin{align}
	\left|\frac{a'}{Ma^2}\right|\ll 1; \quad|\eta|>\eta_a>\eta_c
\end{align}
where, $\eta_a$ is another critical time. This condition enables us to perform WKB analysis for finding the asymptotic solution of \ref{tdhofrw}. The time dependent frequency of the oscillator $\Phi_{\mathbf{k}}$ from \ref{tdhofrw} can now be expanded as
\begin{align}\label{largegammafreq}
\omega_{\mathbf{k}}(\eta)=\frac{a k}{\gamma }+\frac{\gamma  k}{2 a}+\mathcal{O}(\gamma^{2})
\end{align}
Motivated by the correspondence --- between electric field and expanding universe backgrounds --- that we discussed above and the non-perturbative analysis of the electric field case in Ref. \cite{Rajeev2018}, we will assume the following asymptotic behaviour for the scale factor for $|\eta|\gg\eta_c$.
\begin{align}\label{acon1}
&(i)\qquad	a(\eta)\sim\sum_{n=0}^{N}\mathcal{C}_{n}|\mathcal{H}\eta|^{2n-1}\\\label{acon2}
&(ii)\qquad\frac{1}{a(\eta)}\sim\sum_{n=-(N-1)}^{\infty}\tilde{\mathcal{C}}_n|\mathcal{H}\eta|^{2n-1}
\end{align}
for some positive integer $N$. (These correspond to the conditions  Eq(35) and Eq(36) in Ref. \cite{Rajeev2018}.) The positive frequency modes of the asymptotic past ($\phi_{\mathbf{k}({\rm in})}$) and future ($\phi_{\mathbf{k}({\rm out})}$), in the WKB approximation, can then be written as
\begin{align}
	\phi_{\mathbf{k}({\rm in})}\sim\left(\frac{\gamma}{ak}\right)^{1/2}\exp\left[i\int_{-\eta_0}^{\eta}d\eta'\left(\frac{a(\eta') k}{\gamma }\right)+i\int_{-\eta_0}^{\eta}d\eta'\left(\frac{\gamma  k}{2 a(\eta')}\right)\right]\qquad;\eta\rightarrow-\infty\\
	\phi_{\mathbf{k}({\rm out})}\sim\left(\frac{\gamma}{ak}\right)^{1/2}\exp\left[-i\int_{-\eta_0}^{\eta}d\eta'\left(\frac{a(\eta') k}{\gamma }\right)-i\int_{-\eta_0}^{\eta}d\eta'\left(\frac{\gamma  k}{2 a(\eta')}\right)\right]\qquad;\eta\rightarrow\infty
\end{align}  
where, we have assumed that $a(\eta)>0$ for $\eta\gg\eta_c$. Let us use \ref{acon1} and \ref{acon2} to simplify the argument of exponential factors in $\phi_{\mathbf{k}({\rm in})}$ and $\phi_{\mathbf{k}({\rm out})}$, yielding,
\begin{align}
	\int_{-\eta_0}^{\eta}d\eta'\left(\frac{a(\eta') k}{\gamma }\right)\sim\frac{k\log(\mathcal{H}\eta)}{\mathcal{H}\gamma}+\left(\frac{k}{\mathcal{H}\gamma}\right)\sum_{n\neq 0}\frac{\mathcal{C}_n(\mathcal{H}\eta)^{2n}}{2n};\qquad\eta>0\\
	\int_{-\eta_0}^{\eta}d\eta'\left(\frac{\gamma k}{2 a(\eta') }\right)\sim\frac{\gamma k\log(\mathcal{H}\eta)}{2\mathcal{H}}+\left(\frac{\gamma k}{2\mathcal{H}}\right)\sum_{n\neq 0}\frac{\tilde{\mathcal{C}}_n(\mathcal{H}\eta)^{2n}}{2n};\qquad\eta>0
\end{align}
Subsequently we can use these expressions to rewrite $\phi_{\mathbf{k}(in)}$ and $\phi_{\mathbf{k}(out)}$ as
\begin{align}\label{asymexprin}
\phi_{\mathbf{k}({\rm in})}&\sim\left(\frac{\gamma}{|a(\eta)|k}\right)^{1/2}\exp\left[i\left(\left\{\frac{k\mathcal{C}_0}{\mathcal{H}\gamma}+\frac{\gamma k\tilde{\mathcal{C}}_0}{2\mathcal{H}}\right\}\log(-\mathcal{H}\eta)+\sum_{n\neq 0}\left\{\frac{k\mathcal{C}_n}{\mathcal{H}\gamma}+\frac{\gamma \tilde{\mathcal{C}}_nk}{2\mathcal{H}}\right\}\frac{(\mathcal{H}\eta)^{2n}}{2n}\right)\right]\,;\eta\rightarrow-\infty
\\
\label{asymexprout}
\phi_{\mathbf{k}({\rm out})}&\sim\left(\frac{\gamma}{a(\eta)k}\right)^{1/2}\exp\left[-i\left(\left\{\frac{k\mathcal{C}_0}{\mathcal{H}\gamma}+\frac{\gamma k\tilde{\mathcal{C}}_0}{2\mathcal{H}}\right\}\log(\mathcal{H}\eta)+\sum_{n\neq 0}\left\{\frac{k\mathcal{C}_n}{\mathcal{H}\gamma}+\frac{\gamma \tilde{\mathcal{C}}_nk}{2\mathcal{H}}\right\}\frac{(\mathcal{H}\eta)^{2n}}{2n}\right)\right]\,;\eta\rightarrow\infty
\end{align}  
Since, both $\{\phi_{\mathbf{k}({\rm in})},\phi_{\mathbf{k}({\rm in})}^*\}$ and $\{\phi_{\mathbf{k}({\rm out})},\phi_{\mathbf{k}({\rm out})}^*\}$ are a set of linearly independent solutions of \ref{tdhofrw}, we can expand $\phi_{\mathbf{k}({\rm in})}$ in terms of $\{\phi_{\mathbf{k}({\rm out})},\phi_{\mathbf{k}({\rm out})}^*\}$.
\begin{align}\label{bagoluibov}
	\phi_{\mathbf{k}(in)}=\mathcal{A}_{\mathbf{k}}\phi_{\mathbf{k}(out)}+\mathcal{B}_{\mathbf{k}}\phi_{\mathbf{k}(out)}^*
\end{align}
where $\mathcal{A}_{\mathbf{k}}$ and $\mathcal{B}_{\mathbf{k}}$ are the Bogoliubov coefficients. To determine the particle production rate we need to evaluate $\mathcal{B}_{\mathbf{k}}$.  

We will now find an approximate expression for $\mathcal{B}_{\mathbf{k}}$ using the asymptotic expressions for the `in' and `out' modes. This can be done by a procedure, originally due to Landau, which we will call  the Landau procedure. (This was used earlier in \cite{Rajeev2018} in the case of time-dependent electric field, wherein more details can be found. We will not repeat the technical details here.) To use Landau procedure, we will interpret $\eta$ as a complex variable in \ref{asymexprin}. In essence the procedure amounts to rotating $\eta$ in the complex plane from $\arg[\eta]=0$ to $\arg[\eta]=\pi$. We can see that under this transformation the asymptotic expression for $\phi_{\mathbf{k}({\rm in})}$ near $\eta\rightarrow-\infty$  transforms to that of $\phi_{\mathbf{k}({\rm out})}^*$ near $\eta\rightarrow\infty$, except for a constant factor. In view of \ref{bagoluibov}, we can immediately interpret this factor as the Bogoliubov coefficient $\mathcal{B}_{\mathbf{k}}$, which reads,
\begin{align}
	\mathcal{B}_{\mathbf{k}}\approx e^{i\pi}\exp\left[-\pi\left(\frac{k\mathcal{C}_0}{\mathcal{H}\gamma}+\frac{\gamma k\tilde{\mathcal{C}_0}}{2\mathcal{H}}\right)\right]
\end{align}
The number of particles produced can then be computed in a straightforward manner as:
\begin{align}\label{particlelandauds}
n_{\mathbf{k}}&=|\mathcal{B}_{\mathbf{k}}^2|=\exp\left[-2\pi\left(\frac{k\mathcal{C}_0}{\mathcal{H}\gamma}+\frac{\gamma k\tilde{\mathcal{C}_0}}{2\mathcal{H}}\right)\right]\\
\end{align}
Thus, to the leading order, the non-analytic dependence of particle production rate is controlled by the two constants: $\mathcal{C}_0$ and 
$\tilde{\mathcal{C}_0}$. We will now illustrate this result with two examples discussed earlier.

\textit{Example 1:} As a first example of this procedure, consider the physically important case of a locally de Sitter metric, with the scale factor, in some appropriate interval being given by
\begin{align}
a(|\eta|)=\frac{1}{1+H|\eta|}&&\frac{1}{a(\eta)}=1+H|\eta|
\end{align}
so that the relevant constants appearing in the expression for particle number, taking a cue from \ref{acon1} and \ref{acon2}, yields, $\mathcal{C}_{0}=1$, $\tilde{C}_{0}=0$ and $\mathcal{H}\rightarrow H$. Thus the number of particles produced, according to \ref{particlelandauds}, is given by
\begin{align}
	n_{\mathbf{k}}\approx\exp\left[-\frac{2\pi M}{H}\right]
\end{align}
which, we will see in the next section, is consistent with the large mass limit of the exact value of $n_{\mathbf{k}}$. If we treat $M$ as the energy, this is just the Boltzmann limit of a Planck spectrum at temperature $T=H/2\pi$. 

\textit{Example 2:} As a second example,  consider the scale factor given by \ref{radmetric}. The asymptotic expansion now reads $a(\eta)\approx (|a_{0}\eta|/2)$ and hence the inverse scale factor reads, $(1/a(\eta))\approx(2/|a_{0}\eta|)$. With the identification $a_{0}\rightarrow \mathcal{H}$, the coefficients $\mathcal{C}_{0}$ and $\tilde{\mathcal{C}}_{0}$ in this case are given by $0$ and $2$, respectively. The particle number then becomes,
\begin{align}
	n_{\mathbf{k}}\approx\exp\left[-\frac{2\pi k^2}{Ma_0}\right]
\end{align}
which matches \textit{exactly} with \ref{radparticle}. This explicitly demonstrates the usefulness of this approach. 

This concludes the first part of this work, related to the  correspondence between time dependent electric field and expanding universe. We  next want to study the case of time-dependent electric field in an expanding de Sitter background wherein both these effects will be present. However, before we do that (in \ref{dSplusconstE}), we will first consider the case of particle definition and production in the de Sitter universe itself, in order to clarify/highlight several conceptual and technical features in \ref{puredS}.

\section{Particle production in de Sitter spacetime using different approaches}\label{puredS}

So far we have considered particle production in two separate backgrounds, viz., time-dependent electric field in flat spacetime and expanding Friedmann universe, and established a correspondence between them. Our next aim is to consider a situation when a test electric field is present in an expanding Friedmann background, in particular in a de Sitter universe. This will help us to understand the particle production when both the backgrounds are present. We will see that, the case of electric field in de Sitter introduces certain technical and conceptual issues which needs special care. However, these issues arise because of the de Sitter background and are present even in the case of a pure de Sitter universe. The purpose of this section is to review several aspects of particle production in de Sitter spacetime using different approaches, in order to compare and contrast them. The insights obtained in this section will be useful later on when we add an electric field.

With this motivation, we will review particle production in a de Sitter universe \cite{particleinds}. We will be mainly using two approaches; the first one using the mode functions to compute the exact Bogoliubov coefficients and the second one using instantaneous diagonalization of the Hamiltonian for the time dependent oscillator modes to define particle production rate as a function of time. However, we revisit the latter approach using a new  technique, explored in Ref. \cite{demystifying} (also see \cite{Robles-Perez:2017gsc,Rajeev:2017uwk}), that maps \textit{any} time dependent oscillator to an oscillator with constant frequency.  This will give us a handle on several conceptual issues, especially on the definition of particles which is ambiguous in curved spacetime.
\subsection{Particle production using the Bogoliubov Coefficients}\label{dsbogolsec}

	We start by writing down the de Sitter metric in the conformal flat slicing, which takes the following form,
	\begin{align}\label{metricds}
	ds^2=\frac{1}{(1-H \eta)^2}\left(-d\eta^2+|d\mathbf{x}|^{2}\right);\qquad -\infty<\eta<H^{-1}
	\end{align}
	We will consider particle production due to a massive quantum field living in this background spacetime, whose action is given by
	\begin{align}
	\mathcal{A}=\int d^{4}x\sqrt{-g}\left[-\frac{1}{2}g^{a b}\partial _{a}\Phi \partial _{b}\Phi
	-\frac{1}{2}M^{2}\Phi^{2}\right]~
	\end{align}
	where, $g_{ab}$ corresponds to the metric given in \ref{metricds}. (Note that the above action is different from that in \ref{conformaction}, in that the non-minimal coupling term, $R\Phi^2$ is absent. This is because, we are now considering the minimally coupled scalar field, which --- since $R$ is a constant for de Sitter spacetime --- can be generalized in a straightforward manner to arrive at the results of a scalar field with conformal coupling as well.) Let us introduce the Fourier modes $\Phi_{\mathbf{k}}$ by standard means, in terms of which the action simplifies to that of a bunch of time dependent harmonic oscillators, each labelled by $\mathbf{k}$, whose time dependent frequency and mass are given by
	\begin{align}\label{mass_and_freq_eta}
	\omega_{\mathbf{k}}^2(\eta)=k^2+a^2M^2~;\qquad  m_{\mathbf{k}}(\eta)=a^2(\eta)~.
	\end{align} 
	Hence, the equation of motion satisfied by the Fourier mode functions can then be written as
	\begin{align}\label{eominds}
	\frac{d^2\Phi_{\mathbf{k}}}{d\eta^2}+\frac{2H}{1-H\eta}\frac{d\Phi_{\mathbf{k}}}{d\eta}+\left(k^2+\frac{M^2}{(1-H\eta)^2}\right)\Phi_{\mathbf{k}}=0
	\end{align}
	The solution $\phi_{\mathbf{k}}$ to this equation that corresponds to a positive frequency solution in the asymptotic past is given by 
	\begin{align}
	\phi_{\mathbf{k}({\rm in})}(\eta)=\left(\frac{\pi}{4Ha^{3/2}(\eta)}\right)^{1/2} \textrm{H}_{\nu}^{(1)}\left(\frac{k}{Ha}\right)
	\end{align}
	where,
	\begin{align}
	\nu=\sqrt{\frac{9}{4}-\frac{M^2}{H^2}}.
	\end{align}
	This expression for $\nu$ tells us that the situation can be quite different depending on whether $(M/H)$ is greater than or less than $(3/2)$ and we will see that this is indeed the case. One can verify that $\phi_{\mathbf{k}({\rm in})}$ is indeed the positive frequency solution in the early time by noting that as $a\rightarrow 0$ (or, $\eta\rightarrow-\infty$),
	\begin{align}\label{etainfty}
	\phi_{\mathbf{k}}(\eta)\approx \frac{e^{-ik\eta}}{\sqrt{2ka^2(\eta)}}\approx\frac{e^{-i\int d\eta\omega_{\mathbf{k}}(\eta)}}{\sqrt{2m_{\mathbf{k}}(\eta)\omega_{\mathbf{k}}(\eta)}}~.
	\end{align}
	On the other hand, in the late time limit, i.e., as $\eta\rightarrow -(1/H)$, $\phi_{\mathbf{k}}$ takes the form
\begin{align}
\phi_{\mathbf{k}}(\eta)&\approx -\frac{i \sqrt{H} 2^{\nu -1}}{\sqrt{\pi }}\Gamma (\nu ) a^{\nu -\frac{3}{2}} \left(\frac{k}{H}\right)^{-\nu }
+\frac{\sqrt{\pi } \sqrt{H} 2^{-\nu -1}}{\Gamma (\nu +1)}\{1+i \cot (\pi  \nu )\} a^{-\nu -\frac{3}{2}} \left(\frac{k}{H}\right)^{\nu }
\nonumber
 \\
&=A_{\mathbf{k}} a^{-\nu-3/2}+B_{\mathbf{k}} a^{\nu-3/2}
\label{tp1}
\end{align}   
To determine the particle content at late times we have to somehow interpret these two terms as positive and negative frequency \textit{oscillations}, which, of course, is possible only if they are oscillatory. This, in turn happens when $\nu$ is purely imaginary so that we can write $\nu=i|\nu|$. This corresponds to the situation with $M^2/H^2>9/4$, when we can interpret \ref{tp1} as a linear combination of positive and negative frequency modes in the asymptotic future. (The oscillations are with respect to $\ln a\propto t$, the cosmic time, in the asymptotic future while the oscillations are with respect to conformal time $\eta$ in the asymptotic past.)  In this case, one can read off the Bogoliubov coefficients to be
	\begin{align}
	\alpha_{\mathbf{k}}=\frac{A_{\mathbf{k}}}{\sqrt{|A_{\mathbf{k}}|^2-|B_{\mathbf{k}}|^2}};&&
	\beta_{\mathbf{k}}=\frac{B_{\mathbf{k}}}{\sqrt{|A_{\mathbf{k}}|^2-|B_{\mathbf{k}}|^2}}.
	\end{align}
	The number of particles can then be computed as
	\begin{align}
	n_{\mathbf{k}}=|\beta_{\mathbf{k}}|^2=\frac{1}{e^{2\pi|\nu|}-1}
	\label{tp2}
	\end{align}
	which is a \textit{constant}, independent of $k$. \textit{The form of \ref{tp2} is very misleading}; it is \textit{not} a thermal spectrum in the energy of the particle, except when $M\gg H$. Only in this limit, for $k\ll M$, one can interpret \ref{tp2} as a thermal spectrum of particles with a temperature $H/2\pi$. 

The situation gets worse for $M^2/H^2<9/4$. In this case, there are no solutions to \ref{eominds} that behave as positive/negative frequency \textit{oscillatory} modes near $\eta\approx-1/H$. This can be seen from the asymptotic behaviour  of \ref{eominds} in this limit, which takes the following form
	\begin{align}\label{eomindslim}
	\frac{d^2\Phi_{\mathbf{k}}}{d\eta^2}+\frac{2H}{1-H\eta}\frac{d\Phi_{\mathbf{k}}}{d\eta}+\left(\frac{M^2}{(1-H\eta)^2}\right)\Phi_{\mathbf{k}}\approx0
	\end{align}
	The two linearly independent set of solutions of \ref{eomindslim}, with no restriction on the rage of parameters,  are given by:
	\begin{align}\label{nonoscillatory}
	\Phi^{\pm}_{\mathbf{k}}(\eta)&=(1-H\eta)^{\pm\nu+\frac{3}{2}}=e^{-i\mathcal{E}_{\pm}t}
	\end{align}
	where, 
	\begin{align}
	\mathcal{E}_{\pm}=-\frac{3iH}{2}\pm H\sqrt{\frac{M^2}{H^2}-\frac{9}{4}}
	\end{align}
	and we have once again introduced the cosmic time $t$ defined by $(1-H\eta)=e^{-Ht}$. Once again, we see that the notion of positive and negative frequency \textit{oscillatory} modes makes sense only when $M^2/H^2>9/4$. This implies that, for an arbitrary value of mass $M$ outside this range, we cannot define positive and negative frequency modes in a natural fashion and compute the number of particles produced asymptotically. In the next section we circumvent this situation by resorting to a different prescription for defining particles. 
\subsection{Particle number from constant frequency representation}\label{particleapproach}

In a recent work \cite{demystifying} (also see \cite{Robles-Perez:2017gsc,Rajeev:2017uwk}), a rather simple and elegant mapping was found between an arbitrary time dependent harmonic oscillator and a simple harmonic oscillator of unit mass and time \textit{independent} frequency. It was also suggested that one can define particles in a natural manner using this result. We shall first review this approach here for the sake of completeness. 
	
The classical version of this mapping can be summarized as follows: If a dynamical variable $q$ satisfies the \TDHO\ equation with mass and frequency given by $m(\eta)$ and $\omega(\eta)$, respectively, then one can show that the variable $Q=q/f(\eta)$ satisfies the equation of motion of a \textit{constant frequency} oscillator: 
	\begin{align}
		\frac{dQ}{d\tau^2}+\Omega^2Q=0
	\end{align} 
where, $\Omega$ is a constant and we have introduced a new time coordinate $\tau$ through $mf^2d\tau=d\eta$, provided	
the function	$f$ is chosen to be a solution to  the differential equation:
	\begin{align}\label{ermokov}
	\left\{m(\eta)f'\right\}'+\omega^2(\eta)f=\frac{\Omega}{m(\eta)f^3}.
	\end{align}
	 The quantum mechanical  version of this mapping works in a similar way. Let the wave function $\psi(q,\eta)$, for the dynamical system $q$, satisfy the following time dependent Schr\"{o}dinger equation:
	\begin{align}
		i\partial_{\eta}\psi(q,\eta)=\left[-\frac{1}{2m(\eta)}\partial_{q}^2+\frac{1}{2}m(\eta)\omega^2(\eta)q^2\right]\psi(q,\eta).
	\end{align}
	It can then be shown that the new wavefunction $\Psi(Q,\tau)$, defined by
	\begin{align}\label{Eq_psi_phi}
	\psi(q,\eta)=\frac{1}{\sqrt{f}}\exp\left(im(\eta)\frac{f'}{2f}q^{2}\right)\Psi\left[Q=q/f,\tau(\eta)\right]
	\end{align}
	satisfies, the  Schr\"{o}dinger equation for a particle of unit mass in the potential of a simple harmonic oscillator of \textit{constant }frequency $\Omega$, i.e.,
	\begin{align}
		i\partial_{\tau}\Psi(Q,\tau)=\left[-\frac{1}{2}\partial_{Q}^2+\frac{1}{2}\Omega^2Q^2\right]\Psi(Q,\tau),
	\end{align}
	provided that $f$ satisfies \ref{ermokov}. 
	
	This mapping offers a fresh view of the quantization of a time dependent harmonic oscillator and definition of vacuum and particle states. Recall that implicit time dependence of the system implies that there is, in general, no stable vacuum state for a \TDHO. However, there is a unique vacuum state for the $Q$ system, whose wavefunction is given by
	\begin{align}
	\Psi_{0}(Q,\tau)=\left(\frac{\Omega}{\pi}\right)^{1/4}e^{-\frac{\Omega Q^2}{2}}e^{-i\frac{1}{2}\Omega\tau}.
	\end{align}
	Clearly, being an eigenstate of the Hamiltonian of $Q$ system, this state is stationary and hence, once the system is prepared in this state, it continues to be in this state forever. On the other hand, from \ref{Eq_psi_phi}, it follows that, this vacuum corresponds to a time dependent state of the $q$ system whose wavefunction is given by
	\begin{align}
	\psi_0(q;\eta)&=\left(\frac{\Omega}{f^2\pi}\right)^{1/4}\exp\left[-\left(\frac{\Omega}{2f^2}-im\frac{f'}{2f}\right)q^2-\frac{i}{2}\Omega\tau(\eta)\right].
	\end{align}  
	We can expand this state in terms of the complete set of eigenstates, denoted by $\{\phi_{n}(\eta);n=0,1,2...\}$, of instantaneous Hamiltonian of the $q$ system at the instant $\eta$. When the oscillator $q$ corresponds to a time dependent mode function of a physical field in an external background, the average value of the `excitation' parameter $n$, serves as a natural definition for the average number of particles $\bar{n}(\eta)$ produced in that particular mode. A straightforward computation gives (see \ref{constant_freq_mapping} for details),
	\begin{align}\label{averagen}
		\bar{n}(\eta)=\frac{mf^{2}\omega}{4
			\Omega}\left[\left(-1+\frac{\Omega}{mf^{2}\omega}\right)^2+\left(\frac{f'}{f\omega}\right)^2\right]
	\end{align}
		
	As it stands, there is certain degree of arbitrariness in our definition of $\bar{n}$: (i) the constant frequency $\Omega$ can have any value of our choice and we need to fix it by some prescription and (ii) even with a given $\Omega$, since \ref{ermokov} is a second order differential equation, it has a two-parameter infinity  of solutions for the function $f(\eta)$. These ambiguities can be handled as follows: When the time evolution of the system is investigated for the duration $\eta\in(\eta_i,\eta_f)$, a natural choice for $\Omega$ is the value of time dependent frequency $\omega(\eta)$ evaluated at the `early time' $\eta_i$, namely, $\Omega=\omega_i\equiv \omega(\eta_i)$. Further, it is reasonable to demand that the average excitation of the state $\psi_0$  at the early time $\eta_i$ is as small as it can be, if we are to interpret this state as a `vacuum'. With $\Omega=\omega_i$, it follows from \ref{averagen} that, this condition implies
	\begin{align}\label{conditions}
		(i)\quad f(\eta)\approx \sqrt{\frac{\omega_i}{m(\eta)\omega(\eta)}};&&(ii)\quad \frac{f'(\eta)}{f(\eta)\omega(\eta)}\approx 0;\qquad \eta\approx\eta_i.
	\end{align} 
	The combination of conditions (i) and (ii) implies the following `adiabaticity' condition,
	\begin{align}\label{adiabaticcond}
		\Bigg|\frac{1}{2(m\omega^2)}\frac{d(m\omega)}{d\eta}\Bigg|\approx 0;\qquad\eta\approx\eta_i.
	\end{align}
	The vacuum state $\psi_0$, with $\Omega=\omega_i$ and the function $f$ satisfying \ref{conditions}, is equivalent to the adiabatic `in'-vacuum. 
	
	In a similar fashion, we can also define an `out'-vacuum for the $q$ system. To do that, let us first define $\tilde{Q}=q/\tilde{f}$, where $\tilde{f}$ satisfies the differential equation \ref{ermokov} with $\Omega=\omega(\eta_f)\equiv\omega_f$. With $d\tilde{\tau}=dt/(m\tilde{f}^2)$, it then follows that
	\begin{align}
		\frac{d^2\tilde{Q}}{d\tilde{\tau}^2}+\omega_f^2\tilde{Q}=0.
	\end{align}
	We can define an adiabatic `out'-vacuum in exactly the same way we defined the adiabatic `in'-vacuum starting from $\psi_0$. The steps may be summarized as follows: (i) define a state $\tilde{\psi}_{0}$ by replacing $f$ with $\tilde{f}$ (ii) choose $\Omega=\omega_f$ and (iii) demand that $\tilde{f}$ satisfy the conditions in  \ref{conditions}, but at $\eta_f$ instead of $\eta_i$. 
	
	In order to compare this with the standard approach to particle production using Bogoliubov transformations, let us look at the Heisenberg picture version of the formalism discussed above. Recall that the $Q$ and $\tilde{Q}$ oscillators are constant frequency oscillators. Hence, we can define creation and annihilation operators, $\{A,A^{\dagger}\}$ and $\{\tilde{A},\tilde{A}^{\dagger}\}$, respectively, for $Q$ and $\tilde{Q}$ oscillators. In terms of these operators, we can expand the Heisenberg operators $Q$ and $\tilde{Q}$ as 
	\begin{align}
		Q(\tau)=A \frac{e^{-i\omega_i\tau}}{\sqrt{2\omega_i}}+A^{\dagger}\frac{e^{i\omega_i\tau}}{\sqrt{2\omega_i}};&&\tilde{Q}(\tilde{\tau})=\tilde{A} \frac{e^{-i\omega_f\tilde{\tau}}}{\sqrt{2\omega_f}}+\tilde{A}^{\dagger}\frac{e^{i\omega_f\tilde{\tau}}}{\sqrt{2\omega_f}}.
	\end{align}
	Using $q=fQ=\tilde{f}\tilde{Q}$, we get the following expression for the Heisenberg operator $q$:
	\begin{align}
		q&=A\frac{\xi_{{\rm in}}(\eta)}{\sqrt{2\omega_i}}+A^{\dagger}\frac{\xi_{{\rm in}}^*(\eta)}{\sqrt{2\omega_i}}
		=\tilde{A}\frac{\xi_{\rm out}(\eta)}{\sqrt{2\omega_f}}+\tilde{A}^{\dagger}\frac{\xi_{\rm out}^*(\eta)}{\sqrt{2\omega_f}}
	\end{align}
	where, $\xi_{\rm in}(\eta)=f e^{-i\omega_i\tau(\eta)}$ and $\xi_{\rm out}(\eta)=\tilde{f}e^{-i\omega_f\tilde{\tau}(\eta)}$. It is easy to verify that both the sets $\{\xi_{\rm in},\xi_{\rm in}^*\}$ and $\{\xi_{\rm out},\xi_{\rm out}^*\}$ are linearly independent sets of solutions to the \TDHO\ equation of the $q$ system. Hence, we can write
	\begin{align}\label{inasout}
		\xi_{\rm out}=\begin{cases}
		f e^{-i\omega_i\tau}\\
		\alpha \tilde{f} e^{-i\omega_f\tilde{\tau}}+\beta \tilde{f} e^{i\omega_f\tilde{\tau}}
		\end{cases}
	\end{align}
	 where, the Bogoliubov coefficients $\alpha$ and $\beta$ can be found to be 
	 \begin{align}\label{bogol}
	 \alpha&=\frac{\sqrt{m(\eta_f)f^{2}(\eta_f)}}{2\sqrt{\omega_f\omega_i}}\left(\omega_f+\frac{\omega_i}{m(\eta_f)f^{2}(\eta_f)}-i\frac{f'(\eta_f)}{f(\eta_f)}\right)\\
	 \beta&=\frac{\sqrt{m(\eta_f)f^{2}(\eta_f)}}{2\sqrt{\omega_f\omega_i}}\left(-\omega_f+\frac{\omega_i}{m(\eta_f)f^{2}(\eta_f)}-i\frac{f'(\eta_f)}{f(\eta_f)}\right).
	 \end{align}
	The `in'-vacuum $\ket{\rm in}$ can then be defined as the state annihilated by $A$ and similarly, the `out'-vacuum $\ket{\rm out}$ can be defined as the state annihilated by $\tilde{A}$. Then, the number of `out'-particles in the $\ket{\rm in}$ can be evaluated as
	 \begin{align}\label{adiabaticpart}
	 	\bar{n}\equiv \braket{\rm in|\tilde{A}^{\dagger}\tilde{A}|\rm in}=|\beta|^2
	 \end{align}
	 When we are interested in a definition for instantaneous particle production rate, we can always choose the time $\eta_f$ to be at the desired instant of time, say $\eta$. In this case, the Bogoluibov coefficients become a function of this time $\eta$ and hence, the number of particles $\bar{n}$, which we will denote as $\bar{n}(\eta)$.
	 
	  It is easy to verify that $\bar{n}(\eta)$ evaluated in this manner using the expression for $\beta$ in \ref{bogol} matches exactly with that in \ref{averagen}. However, it is worth mentioning that there is a subtle difference in interpretation. The derivation of \ref{adiabaticpart} assumes the existence of an adiabatic `out'-vacuum, which in turn depends on the validity of adiabatic condition \ref{adiabaticcond}, but at $\eta_f$ instead of $\eta_i$. This condition, in general, may not hold at $\eta_f$ and hence, the interpretation of $\bar{n}$ as the number of `out'-vacuum makes sense only when this condition holds. Moreover, when \ref{adiabaticcond} is valid, the standard approach to particle production bases on the Bogoliubov transformation connecting the asymptotic `in' and `out' modes and the approach that we discussed bases on $\{A,A^{\dagger}\}$ and $\{\tilde{A},\tilde{A}^{\dagger}\}$ are equivalent. On the other hand, the derivation of \ref{averagen} was based on the eigenstates of instantaneous Hamiltonian the $q$ oscillator is always well defined as long as $\omega^2$ and $m$ are positive.
\subsection{Application to particle production in de Sitter}

In this section we will compute the particle number associated with a certain Fourier mode of a scalar field in de Sitter spacetime, using the  formalism presented above. This uses the fact that the Fourier modes of a scalar field in de Sitter spacetime can be transformed into a time dependent harmonic oscillator, whose frequency and mass takes the following form,
	\begin{align}\label{timedepfre}
	\omega_{\mathbf{k}}^2(\eta)&=k^2+M^2a^2\\
	m(\eta)&=a^2(\eta)
	\end{align}
Using the redefinition of the dynamical variable through the function $f_{\mathbf{k}}(\eta)$ one can convert the time dependent oscillator to a constant frequency oscillator with unit mass. The frequency of the constant frequency oscillator has been fixed to be $\omega_{\mathbf{k}(i)}$ associated with some initial time $\eta=\eta_{i}$. Thus the time dependent Bogoliubov coefficients associated with particle production at late times take the form,
	\begin{align}\label{alpha}
	\alpha_{\mathbf{k}}&=\frac{1}{2}\sqrt{\frac{mf_{\mathbf{k}}^2}{\omega_{\mathbf{k}}\omega_{\mathbf{k}(i)}}}\left[\omega_{\mathbf{k}}+\frac{f_{\mathbf{k}}\omega_{\mathbf{k}(i)}}{mf_{\mathbf{k}}^2}-i\frac{f_{\mathbf{k}}'}{f_{\mathbf{k}}}\right]\\\label{beta}
	\beta_{\mathbf{k}}&=\frac{1}{2}\sqrt{\frac{mf_{\mathbf{k}}^2}{\omega_{\mathbf{k}}\omega_{\mathbf{k}(i)}}}\left[-\omega_{\mathbf{k}}+\frac{f_{\mathbf{k}}\omega_{\mathbf{k}(i)}}{mf_{\mathbf{k}}^2}-i\frac{f_{\mathbf{k}}'}{f_{\mathbf{k}}}\right]
	\end{align}
Here the unknown function $f_{\mathbf{k}}$ connecting the Fourier modes of the time dependent oscillator to constant frequency oscillator satisfies the following differential equation
	\begin{align}\label{ermokovpinney}
	\left\{m(\eta)f_{\mathbf{k}}'\right\}'+\omega_{\mathbf{k}}^2(\eta)f_{\mathbf{k}}=\frac{\omega_{\mathbf{k}(i)}^2}{m(\eta)f_{\mathbf{k}}^3}
	\end{align}
Two comments are in order at this stage. First, in the case of a \textit{single} \TDHO, we have a \textit{single} scaling function $f$ (and corresponding transformation to the new time coordinate $\tau$) and everything proceeds smoothly. When we apply this idea to QFT in Friedmann universe, we have a \textit{set} of $f_{\mathbf{k}}$ which depend on $\mathbf{k}$. This, in turn, means that the new time coordinate $\tau_{\mathbf{k}}$ will also be different for each mode. Treated as a trick to solve the equations of motion (in Heisenberg picture) or the Schr\"{o}dinger equation (in Schr\"{o}dinger picture) this creates no problem. But it is \textit{not} a global coordinate transformation of the time coordinate in the spacetime.

Second, let us comment on the choice of the initial vacuum obtained by this approach, compared to the standard one. As pointed out earlier, the frequency of the constant frequency oscillator is taken to be $\omega_{\mathbf{k}(i)}\equiv \omega_{\mathbf{k}}(\eta_i)$. In the context of de Sitter spacetime all the initial information is usually specified at the asymptotic past:  $\eta_i\rightarrow-\infty$. As evident from \ref{timedepfre}, it follows that, in this limit $\omega_{\mathbf{k}(i)}=k$ and the function $f_{\mathbf{k}}$ becomes, 
	\begin{align}
	f_{\mathbf{k}}=\left(\frac{k\pi}{2Ha^{3}(\eta)}\right)^{1/2}\left|\textrm{H}_{\nu}^{(1)}\left(\frac{k}{H a(\eta)}\right)\right|\sim \frac{1}{a(\eta)},
	\end{align}
	which, in the asymptotic past behave as $f_{\mathbf{k}}\sim1/a(\eta)$. Since, in the case of de Sitter spacetime the mass of the scalar field scales as $a^2$ and $\omega_{\mathbf{k}}(\eta)\approx k$, in the $\eta\rightarrow -\infty$ limit the function $f_{\mathbf{k}}$ behaves as $f_{\mathbf{k}}\approx(m(\eta)\omega_{\mathbf{k}}(\eta)\omega_{\mathbf{k}}^{-1}(-\infty))^{-1/2}$. This is in exact agreement with our discussion in the previous section. Further, the corresponding `in-mode' function matches exactly with that associated to the Bunch-Davies vacuum state. \textit{Hence the vacuum state defined in the context of constant frequency oscillator actually corresponds to the standard Bunch-Davies vacuum.} So the particle number at later times, calculated by this approach, should also correspond to the standard situation when the quantum field starts from the Bunch-Davies vacuum. Finally, let us examine the validity of the adiabatic condition at $\eta_i=-\infty$ and $\eta_f=H^{-1}$.
	\begin{align}
		\Bigg|\frac{1}{2\left[m(\eta)\omega_{\mathbf{k}}^2\right]}\frac{d}{d\eta}\left[m(\eta)\omega_{\mathbf{k}}\right]\Bigg|\approx\begin{cases}
		|k\eta|^{-1}&;\qquad\eta\rightarrow-\infty\\
		\dfrac{3H}{2M}&;\qquad\eta\approx H^{-1}.
		\end{cases}
	\end{align}
	The adiabatic condition clearly holds well in the early times. On the other hand, in the asymptotic late times the adiabatic condition holds only for $M/H\gg3/2$, hence, the interpretation of $|\beta_{\mathbf{k}}(H^{-1})|^2$ as the number of `out'-particles makes sense only in this limit. This is exactly the reason why we encountered a problem in \ref{dsbogolsec} while trying to study particles production for $M/H<3/2$ using the conventional approach. However, for any range of $M/H$ we can always find	$\beta_{\mathbf{k}}(\eta)$ using \ref{beta} and compute the number of particles as $|\beta_{\mathbf{k}}(\eta)|^2$, which has to be interpreted in terms of the eigenstates of instantaneous Hamiltonian of the \TDHO\ corresponding to the Fourier modes. In what follows, we will study the particle production using the approach discussed in \ref{particleapproach} for all values of $M/H$.  
	\subsubsection{Massless fields in de Sitter}

We saw earlier that the standard approach ran into trouble when $M/H<3/2$ because the modes associated with the field are non-oscillatory in the asymptotic future, thereby making the definition of particles problematic.   To show that the approach based on constant frequency mapping works in this case, we shall first work out the massless case explicitly. In the constant frequency approach, one does not encounter any hurdles and it is possible to write down the particle number using the following solution for the function $f_{\mathbf{k}}(\eta)$, 
	\begin{align}
	f_{\mathbf{k}}(\eta)=\left(\frac{1}{2a(\eta)^2k}+\frac{H^2}{2k^3}\right)^{1/2}
	\end{align}
	Given the function $f_{\mathbf{k}}(\eta)$, one can immediately determine the Bogoliubov coefficients and hence the particle number from the relation $n_{\mathbf{k}}=|\beta_{\mathbf{k}}|^2 $. This leads to the following expression for the particle number,
	\begin{align}\label{particlemassless}
	n_{\mathbf{k}}=\frac{H^2}{4 k^2 (1-H\eta)^2}=\frac{H^2a(\eta)^2}{4k^2}
	\end{align}
	where, we have used the fact that scale factor for de Sitter spacetime behaves as $(1-H\eta)^{-1}$. We see that $n_{\mathbf{k}}$ diverges in the late time limit as $a\rightarrow\infty$ (or, equivalently as $\eta\rightarrow (1/H)$). (We  note that a similar divergence was also noticed earlier in \cite{Mahajan2008}.) However, it is interesting to note that the proper number density of particles with \textit{physical} momentum $\mathbf{p}=\mathbf{k}/a$ inside a spherical shell in $\mathbf{p}$-space of radius $p$ and thickness $dp$, is finite, constant and independent of $p$.
	\begin{align}\label{masslessN}
		n_{\mathbf{k(\mathbf{p})}}\frac{4\pi p^2dp}{(2\pi)^3}&=\left(\frac{H^2}{8\pi^2}\right)dp
	\end{align}
	We shall next consider the massive field.
	\subsubsection{Massive field in de Sitter satisfying $(M^2/H^2)<(9/4)$}

In this case as well the standard mode function technique falls silent about particle definition at late times, again due to non-oscillatory behaviour of the mode function. On the other hand, the constant frequency technique indeed yields an expression for the particle number. For this purpose, we need to first determine the form of $f_{\mathbf{k}}$, which is now given by:
	\begin{align}\label{f_for_general}
	f_{\mathbf{k}}=\left(\frac{k\pi}{2Ha^{3}(\eta)}\right)^{1/2}\left[J_{\nu}^2\left(\frac{k}{aH}\right)+Y_{\nu}^2\left(\frac{k}{a H}\right)\right]^{1/2},
	\end{align}
	Given this $f_{\mathbf{k}}$ one can immediately find out the Bogoliubov coefficients using \ref{bogol} and compute the instantaneous particle number using the  relation, $n_{\mathbf{k}}=|\beta_{\mathbf{k}}|^2$. In the general context, the expression for $n_{\mathbf{k}}$ is complicated and not very enlightening.  however the  asymptotic expression for the particle number is simpler and we will concentrate on that case. It is useful to start with the  the asymptotic form of  $n_{\mathbf{k}}$ as $a\rightarrow\infty$. In this limit $f_{\mathbf{k}}$ behaves as
	\begin{align}
	f_{\mathbf{k}}\approx\frac{2^{\nu -\frac{1}{2}}}{\sqrt{\pi }} a^{\nu -\frac{3}{2}} \Gamma (\nu ) \left(\frac{k}{H}\right)^{\frac{1}{2}-\nu }
	\end{align}
	and hence the asymptotic particle number takes the following form,
	\begin{align}\label{massless}
	n_{\mathbf{k}}\approx a^{2 \nu }\left[\frac{2^{2 \nu -5}  \Gamma (\nu )^2 \left(\frac{k}{H}\right)^{-2 \nu } \left\{H^2 (3-2 \nu )^2+4 M^2\right\}}{(\pi  H M)}\right]
	\end{align}
	
	Thus, in this case as well we find that $n_{\mathbf{k}}\rightarrow\infty$ as $a\rightarrow\infty$. However, there is a bit of subtlety in the massless case which we will comment on. We know from the earlier discussion (see \ref{particlemassless}) that, in the massless limit, corresponding to $\nu\rightarrow (3/2)$, the particle number should vary as $a^2$. But, if we naively take the $\nu\rightarrow (3/2)$ limit of \ref{massless} the particle number seems to vary as $\sim a^{3}$ rather than as $\sim a^{2}$. This arises because the intermediate steps in the calculation  involve handling the combination $ M^2a^2$ and its limiting value depends on the order in which the limits $a\to\infty$ and $M\to0$ are taken. If one takes $M\to0$ limit first --- at finite $a$ --- the combination $Ma$ reduces to zero; but if one first takes $a\to\infty$ --- with nonzero $M$ ---  the combination $Ma$ diverges. This requires us to be careful in defining the two limits. We can see this more clearly by defining a function $\mathcal{N}_{\mathbf{k}}(\nu,a)$ given by
	\begin{align}
		\mathcal{N}_{\mathbf{k}}(\nu,a)\equiv n_{\mathbf{k}}a^{-2\nu}.
	\end{align}
	Let us now consider the following two limits of this functions: (i) $\nu\rightarrow 3/2$ followed by $a\rightarrow\infty$ and (ii)  $a\rightarrow\infty$ followed by $\nu\rightarrow 3/2$. We find that:
	\begin{align}\label{nu_then_a}
		&\textrm{(i)} \lim_{a\rightarrow\infty}\lim_{\nu\rightarrow 3/2}\mathcal{N}_{\mathbf{k}}(\nu,a)=0\\\label{a_then_nu}
		&\textrm{(ii)}\lim_{\nu\rightarrow 3/2}\lim_{a\rightarrow\infty}\mathcal{N}_{\mathbf{k}}(\nu,a)=\infty
	\end{align}
	More details can be found in \ref{disc}.
\subsubsection{Massive fields in de Sitter satisfying $(M^2/H^2)>(9/4)$}

	In this final section, we will consider the case of massive fields in de Sitter satisfying $(M^2/H^2)>(9/4)$ for which the particle interpretation had no ambiguity, even in the standard approach. In this case, $\nu=i|\nu|$ and we have the following expression for the function $f_{\mathbf{k}}$, relating modes of the time dependent oscillator to those of the constant frequency oscillator. 
	\begin{align}
	f_{\mathbf{k}}^2=\left(\frac{\pi  H^2 x^3 }{2 k^2}\right)\textrm{H}_{i |\nu| }^{(1)}(x) \textrm{H}_{i |\nu| }^{(2)}(x)
	\end{align}
where, $x=k/[Ha(\eta)]$. Since, we are interested in the particle production in the late times, let us look at the $x\ll 1$ limit of $f_{\mathbf{k}}^2$, which is given by:
\begin{align}
	f_{\mathbf{k}}^2\approx\frac{H^2 x^3 \text{csch}(\pi  |\nu| )}{k^2 |\nu| }\left[\cosh (\pi  |\nu| )+\cos (2 |\nu|  \log (x))\right]
\end{align}
	Using the above approximation for  $f_{\mathbf{k}}$, we can find the following leading order expressions for different factors needed to evaluate $|\beta_{\mathbf{k}}|^2$.
\begin{align}
	&\quad\frac{m(\eta)f_{\mathbf{k}}^2}{4k\omega_{\mathbf{k}}(\eta)}\approx\frac{H^2 x^2 \text{cot}(\pi  |\nu| )}{4Mk^2 |\nu| }\\
	&\quad\left(-\omega_{\mathbf{k}}(\eta)+\frac{\omega_{\mathbf{k}}}{m(\eta)f_{\mathbf{k}}^{2}}\right)^2\approx \frac{k^2 }{x^2}\left(\frac{M}{H}-\frac{|\nu|  \sinh (\pi  |\nu| )}{\cosh (\pi  |\nu| )+\cos (2 |\nu|  \log (x))}\right)^2\\
	&\quad \left(\frac{f_{\mathbf{k}}'}{f_{\mathbf{k}}}\right)^2\approx\frac{\left[3 k \left\{\cosh (\pi  |\nu| )+\cos (2 |\nu|  \log x)\right\}-2 k |\nu|  \sin (2 |\nu|  \log x)\right]^2}{4 x^2 \left[\cosh (\pi  |\nu| )+\cos (2 |\nu|  \log x)\right]^2}
\end{align}
The number density of particles can now be calculated using \ref{averagen} and found to be
\begin{align}
	n_{\mathbf{k}}\approx \frac{-8 |\nu| ^2+\left(8 |\nu| ^2+9\right) \coth (\pi  |\nu| )+3 \text{csch}(\pi  |\nu| ) (3 \cos (2 |\nu|  \log x)-4 |\nu|  \sin (2 |\nu|  \log x))}{16 |\nu|^2}
\end{align}
The cosine and sine terms in this expression, as $x\rightarrow 0$, oscillates fast and averages to zero. Thus, the above expression simplifies to
\begin{align}\label{particleds}
	n_{\mathbf{k}}\approx\frac{\left(8 |\nu| ^2+9\right) \coth (\pi  |\nu| )-8 |\nu| ^2}{16 |\nu|^2}
\end{align}
Recall that an adiabatic `out'-vacuum exists only when $M/H\gg3/2$, in which case we can also assume that $|\nu|\gg 1$. In this limit, \ref{particleds} further simplifies to
\begin{align}\label{dsparticle}
	n_{\mathbf{k}}\approx \frac{1}{e^{2\pi|\nu|}-1}
\end{align} 
and matches with the earlier result in \ref{tp2}.

\subsubsection{Brief comparison of the two approaches} 

As was pointed out earlier, the particle number obtained in \ref{averagen} is valid for \textit{all times} as long as $\omega^2$ is a strictly positive function. This is the reason why, in this approach, we could define $n_{\mathbf{k}}$ for all values of $M$. However, we saw in \ref{nonoscillatory} that conventional approach to finding particle production using mode functions seems to fail when there are no asymptotically oscillatory basis of solutions, even if $\omega^2>0$. The non-existence of asymptotically oscillatory solutions is related to the fact that the adiabaticity condition, given by \ref{adiabaticcond} is violated at late times. To understand this, let us again consider a general time-dependent oscillator for which the out-mode can be written as
\begin{align}
	\xi_{out}(\eta)=\tilde{f}e^{-i\omega_f\tilde{\tau}(\eta)}.
\end{align}
Near $\eta=\eta_f$, the right hand side approximates to
\begin{align}
	\xi_{out}(\eta)\approx\sqrt{\frac{\omega_f}{m(\eta)\omega(\eta)}}e^{-i\int d\eta\omega(\eta)}
\end{align}
There are two possible situations which can arise: first is when the phase factor oscillates much faster that the rate of variation of amplitude and the second is when the phase factor oscillates much slower than the rate of variation of the amplitude. Depending on whether the adiabatic condition in \ref{adiabaticcond} holds or not near $\eta_f$, we get the first and second scenarios, respectively. Now, in the de Sitter case, when $M/H\ll3/2$, we saw that the adiabatic condition is violated. Therefore, the asymptotic solutions have effectively constant phases; however, the amplitudes are time dependent. On the other hand, when $M/H\gg3/2$, the oscillation of phase factor becomes relevant, as is reflected in the existence of asymptotic oscillatory solutions. We shall now compare the conventional method with the approach presented in \ref{particleapproach}, keeping these two cases in mind. In particular, we will show that the mode function associated with the constant frequency oscillator at late times coincide with the `out'-modes discussed in \ref{tp1}. 

The time evolution of the `in'-mode in the constant frequency approach is given by \ref{inasout} and hence for a scalar field in de Sitter, the `in' mode can be expressed in the early and late times as,
\begin{align}\label{dsmodeasympt}
	\phi_{\mathbf{k}({\rm in})}=\begin{cases}
	f_{\mathbf{k}}e^{-i\omega_{\mathbf{k}}(\eta_i)\tau_{\mathbf{k}}}\\
	\alpha^{f}_{k}\tilde{f}_{\mathbf{k}}e^{-i\tilde{\omega}_{\mathbf{k}}(\eta_f)\tilde{\tau}_{\mathbf{k}}}+\beta^{f}_{k}\tilde{f}_{\mathbf{k}}e^{i\tilde{\omega}_{\mathbf{k}}(\eta_f)\tilde{\tau}_{\mathbf{k}}}
	\end{cases}
\end{align} 
 where, $\alpha^{f}_{\mathbf{k}}=\alpha_{\mathbf{k}}(H^{-1})$, $\beta^{f}_{\mathbf{k}}=\beta_{\mathbf{k}}(H^{-1})$. Further, $\eta_i\approx-\infty$ refers to the initial time and $\eta_f\approx H^{-1}$ correspond to the final time. Using the limiting behaviour of $f_{\mathbf{k}}$ and $\tilde{f}_{\mathbf{k}}$ given in \ref{conditions}, we can show that the new time coordinates $\tau_{\mathbf{k}}$ and $\tilde{\tau}_{\mathbf{k}}$ have the following limits, 
 \begin{align}\label{newt1}
 	\tau_{\mathbf{k}}&= \int \frac{d\eta}{a^2f_{\mathbf{k}}^2}\approx \eta;\qquad\eta\rightarrow-\infty\\\label{newt2}
\tilde{\tau}_{\mathbf{k}}&=	\int \frac{d\eta}{a^2\tilde{f}_{\mathbf{k}}^2}\approx -\frac{1}{Ha(\eta_f)}\textrm{log}(1-H \eta);\qquad\eta\approx H^{-1}
 \end{align}
Note that, near $\eta=H^{-1}$, the coordinate $\tilde{\tau}_{\mathbf{k}}$ is proportional to the cosmic time $t=-H^{-1}\log(1-H\eta)$. When $M/H\gg3/2$, to the leading order approximation, we have $M/H\approx|\nu|$. Using this fact and the approximate values of the new time coordinates given in \ref{newt1} and \ref{newt2}, we get
\begin{align}
		\phi_{\mathbf{k}({\rm in})}\propto\begin{cases}
	\dfrac{e^{-ik\eta}}{a(\eta)};\qquad \eta\rightarrow-\infty\\
	\alpha^{f}_{k}\dfrac{e^{-iH|\nu| t}}{a^{3/2}(\eta)}+\beta^{f}_{k}\dfrac{e^{iH|\nu|t}}{a^{3/2}(\eta)};\qquad\eta\approx H^{-1}.
	\end{cases}
\end{align}
The last line  is in perfect agreement with the $\eta\approx H^{-1}$ approximation of the `out'-mode function given in \ref{tp1}. Thus starting from the initial mode function in the constant frequency approach, we can relate it to the `in' and `out' mode functions of the conventional approach in the limit $(M/H)\gg (3/2)$. This explicitly demonstrates that the constant frequency approach is consistent with conventional approach in the limit of large $(M/H)$. 

 We conclude with a comment about more general expansion laws. It has been shown in Ref.\cite{Lochan:2018pzs} that  the dynamics of a \textit{massless} field in a universe with a power law expansion, such that $a(\eta)\sim (1-H\eta)^{-q}$, where $0<q<1$, can be mapped to the dynamics of a \textit{massive} field in de Sitter universe with $a\sim(1-H\eta)^{-1}$  and mass, $M^{2}=(1-q)(2+q)H^{2}$. Thus massless particle production in a power law universe can be directly mapped to massive particle production in de Sitter spacetime. However, as evident from \ref{dsbogolsec}, the approach in terms of  mode functions will not work satisfactorily in this case, since $(M^{2}/H^{2})>(9/4)$ will imply $-(-2q-1)^{2}>0$, which can not be satisfied irrespective of the choice of $q$. Thus one must use the mapping to the constant frequency oscillator described in \ref{particleapproach}. In that scenario, as evident from \ref{massless}, the number density of particles scales as $H^{2\nu}$. This vanishes in the $H\rightarrow 0$ limit, as to be  expected, since $a(\eta)$ becomes unity as $H\rightarrow 0$. Moreover, the particle number density vanishes as  an analytic function of the the coupling  parameter.  Note that only when we consider a massive scalar field in de Sitter with $(M/H)>(3/2)$ the non-analytic behaviour of particle number in Hubble parameter kicks in.

\section{Constant electric field in de Sitter}\label{dSplusconstE}

In the following sections we are going to explore pair production when there is both an expanding scale factor as well as a time dependent electric field. However, for the sake of simplicity, we will be mainly concerned with a scale factor that corresponds to the de Sitter universe. 

In the previous section we have explicitly determined the particle number in the context of de Sitter spacetime. There has been some interest in the literature \cite{schwingerinds1,schwingerinds2,schwingerinds3,Sharma:2017ivh} to study the particle number when a constant electric field is present in de Sitter spacetime. Our main aim is to determine the particle number in the context of a time dependent electric field in de Sitter; however, it is useful to  discuss the case of a constant electric field  first. We will work exclusively in the spatially flat Friedmann universe, expressed in terms of the conformal time $\eta$, in which case the scale factor $a(\eta)$ is given by \ref{metricds}.  The constant electric field must be defined in a covariant way and the most natural choice being $F^{\mu\nu}F_{\mu\nu}=\textrm{constant}\equiv-2E^{2}_{0}$. 
We will assume that the field is along the $z$ direction and is described by the (spatial) vector potential, $\mathbf{A}=\{0,0,A_{z}(\eta)\}$. This implies, given the above definition of constant electric field, that $A_{z}(\eta)$ must satisfy the following differential equation, $\partial_{\eta}A_{z}=-a^2E_0$. This equation can be immediately integrated, yielding the following expression for the vector potential,
\begin{align}
A_z=-\int\frac{E_0 d\eta }{(1-H\eta)^2}=-\frac{E_0}{H}\frac{1}{(1-H\eta)}+\textrm{constant}.
\end{align}
The constant of integration must be chosen carefully such that in the $\eta\rightarrow 0$ limit $A_z(\eta)$ turns out to be finite. This fixes the constant to be, $(E_{0}/H)$. With this choice for the constant, the vector potential turns out to be,
\begin{align}
A_z&=-\frac{E_0}{H}\frac{H\eta}{1-H\eta}
\end{align}
Note that, when $H=0$ spacetime becomes flat and this expression reduces to the one in standard flat spacetime Schwinger effect. (If this is not ensured, the final result may not have correct $H\to0$ limit, which has happened in some of the previous discussions in the literature.)

As an aside, we will comment on the source for the electromagnetic field which is usually not stressed in the literature. 
In general, a time dependent electric field gives rise to a magnetic field. If that is the case, we need to take into account the effect of the magnetic field in the computation of particle production. However,  in the present context,  the vector potential (assumed to be along $z$ direction) depends on time alone. Therefore, the only non-trivial component of $F_{\mu \nu}$ corresponds to $F_{0z}=\dot{A}_{z}=-E_{z}$ (say, in flat spacetime) and there are no magnetic fields. But for consistency of Maxwell's equations we \textit{must} have a non-zero current, which is given by $J^{\nu}=(1/4\pi)\nabla _{\nu}F^{\mu \nu}$. If we consider time dependent electric field in a flat spacetime, then it follows that, $J^{z}$ is the only non-zero component, such that, $J^{z}=(1/4\pi)\dot{E}_{z}$. In the cosmological spacetime as well, we have $J^{z}$ to be the only non-zero component, but its explicit form becomes, $J^{z}=(1/4\pi)a^{-4}(\eta)\partial _{\eta}E_{z}$. Hence, in all the cases considered here, there are no magnetic fields but a non-trivial current must exist to ensure that we have a purely electric field situation. (The role of the current is not usually discussed in the literature --- and we will also ignore it --- though it may be worth investigating for a more complete picture.)

We now consider a complex scalar field in this background. Its Fourier modes will satisfy the equation of a time dependent harmonic oscillator, with unit mass and time dependent frequency. This is essentially a generalization of \ref{freqFriedmann}, which, for this background, yields the time dependent frequency to be,
\begin{align}
\omega^2_{\mathbf{k}}=k_{\perp}^2+\frac{M^2}{(1-H\eta)^2}+\left(k_z+\frac{qE_0}{1-H\eta}\right)^2.
\end{align}
It can be easily verified that in the $H\rightarrow 0$ limit, the frequency becomes, $M^2+k_{\perp}^2+(k_z+qE_0\eta)^2$, which is consistent with that of the Schwinger effect discussed in \cite{Rajeev2018}. On the other hand, in the $qE_{0}\rightarrow 0$ limit, it immediately follows that the frequency becomes, $k^2+M^2(1-H\eta)^{-2}$, coinciding with the frequency of $\Phi_{\bf k}$ in the de Sitter spacetime. In what follows we will study the particle production in this background by three different methods, namely, (i) Landau procedure (ii) Hamilton-Jacobi method and (iii) using mode functions. 

\subsection{Landau Procedure}\label{landaudssec}

In \ref{nonpert}, we discussed a procedure to calculate the WKB limit of particle production. We will now apply the same techniques -- the Landau procedure -- to compute particle production for our current case. It turns out  that we do not have to redo the whole calculation in \ref{nonpert} to find the generalization of \ref{particlelandauds} to the present problem. Instead we can proceed as follows: Let us first look at the expansion of $\omega_{\mathbf{k}}$ for large $|\eta|$, which takes the form
	\begin{align}\label{freqexpansdsschwinger1}
	\omega_{\mathbf{k}}&\approx\frac{M}{H\eta}+\frac{H\eta}{2M}\left\{k_{\perp}^2+(k_z-qE_0/H)^2\right\}
	\end{align}
	Comparing this with \ref{acon1}, \ref{acon2} we see the following identification of parameters exist:
	\begin{align}\label{identificaton1}
		\frac{M}{H}\rightleftarrows \left(\frac{\mathcal{C}_0k}{\mathcal{H}\gamma}+\frac{\gamma k\tilde{\mathcal{C}}}{2H}\right)
	\end{align} 
	Hence, using \ref{particlelandauds}, the particle number can be immediately written down as:
	\begin{align}\label{ndslandau}
	n_{\mathbf{k}}=\exp{\left[-\frac{2\pi M}{H}\right]}
	\end{align}
	This is however only the leading order term as can be seen from the fact that this expression is independent of the electric field!
	Thus, the Landau procedure, to the \textit{leading} order only captures a factor independent of $qE_0$.  

	To get the dependence on the electric field and $H$ then we need to retain more terms in the asymptotic expansion. That is, when $\eta\rightarrow \infty$, one must keep $H\eta\rightarrow\textrm{finite}\ll1$. Then the  expansion in \ref{freqexpansdsschwinger1} should be replaced by
	\begin{align}\label{omegaforHzero}
	\omega_{\mathbf{k}}&\approx\sqrt{q^2E_0^2+3M^2H^2}\eta+\frac{M^2H}{\sqrt{q^2E_0^2+3M^2H^2}}\\\nonumber
	&+\frac{1}{2}\left[\frac{k_{\perp}^2+M^2}{\sqrt{q^2E_0^2+3M^2H^2}}-\frac{M^4H^2}{(q^2E_0^2+3M^2H^2)^{3/2}}\right]\frac{1}{\eta}.
	\end{align}
	Further, the identification in \ref{identificaton1} should be replaced by
	\begin{align}
		\frac{1}{2}\left[\frac{k_{\perp}^2+M^2}{\sqrt{q^2E_0^2+3M^2H^2}}-\frac{M^4H^2}{(q^2E_0^2+3M^2H^2)^{3/2}}\right]\equiv \Theta(H,M,qE_0)\rightleftarrows \left(\frac{\mathcal{C}_0k}{\mathcal{H}\gamma}+\frac{\gamma k\tilde{\mathcal{C}}}{2H}\right).
	\end{align}
	Thus, the particle number density evaluates to:
	\begin{align}
		n_{\mathbf{k}}=e^{-2\pi\Theta}.
	\end{align}
	This expression has correct limits. In particular, it can easily be verified that, in the $H\rightarrow 0$ limit,  this expression reduces to
	\begin{align}
	n_{\mathbf{k}}=\exp{\left[-\frac{\pi(k_{\perp}^2+M^2)}{qE_0}\right]}
	\end{align}
	which is the standard result in Schwinger effect.
\subsection{Euclidean action approach}\label{euclideandssec}

Another elegant method of computing semi-classical limit of particle number is by using the Euclidean action. The idea is to first evaluate the action $\mathcal{A}_{E}$ for an appropriate classical solution of the Euclidean equation of motion for a hypothetical particle. It can be shown that, for the cases which are of interest to us, the particle number, when $\mathcal{A}_{E}\gg1$, is given by
\begin{align}\label{neuclideanA}
n\approx \exp{\left(-\mathcal{A}_{E}\right)}.
\end{align}
The Euclidean action is most easily computed by solving the Hamilton-Jacobi (HJ) equation. For our case, let us denote by $\mathcal{A}$ the action for a charged particle in the Friedmann spacetime with the scale factor $a(\eta)$ and constant electric field. The Hamilton-Jacobi equation in this context is given by,
\begin{align}
\frac{1}{a^2}\left[-\partial_{\eta}\mathcal{A}+|\partial_{_{x_{\perp}}}\mathcal{A}|^2+(\partial_{z}\mathcal{A}-qA_{z})^2\right]=-M^2a^2
\end{align}
The symmetry of the problem suggests the ansatz, $A_{z}=k_z z+\mathbf{k}_{\perp}.\mathbf{x}_{\perp}+F(\eta)$, where $F(\eta)$ satisfies
\begin{align}
-\left(\partial_{\eta}F\right)^2+k_{\perp}^2+(k_z-qA_z)^2=-M^2a^2
\end{align}
This can be integrated to give,
\begin{align}
\mathcal{A}=\int d\eta\sqrt{M^2a^2+k_{\perp}^2+(k_z-qA_z)^2}+k_z z+\mathbf{k}_{\perp}.\mathbf{x}_{\perp}
\end{align}
In particular, for the de Sitter spacetime, the classical action evaluates to
\begin{align}\label{actionconstEds}
\mathcal{A}=\int \frac{d\eta}{1-H\eta}\sqrt{M^2+k_{\perp}^2(1-H\eta)^2+\left\{qE_0\eta+k_z(1-H\eta)\right\}^2}+k_z z+\mathbf{k}_{\perp}.\mathbf{x}_{\perp}
\end{align}
	
We are interested in closed classical trajectories (in the Euclidean sector), for which the last two terms in \ref{actionconstEds} vanish. Then, a straightforward computation shows that there are two complex turning points, defined by the vanishing of square root terms in the integrand in \ref{actionconstEds}. These turning points are given by
\begin{align}
H\eta_{\pm}=\frac{k_{\perp}^2+k_z(k_z-qE_0H^{-1})\pm i M\sqrt{k_{\perp}^2+(k_z-qE_{0}H^{-1})^2+(qE_{0})^2M^{-2}H^{-2}}}{k_{\perp}^2+(k_z-qE_{0}H^{-1})^2}.
\end{align}
The expression for number of particles give in \ref{neuclideanA} is good approximation only for sufficiently large value of the Euclidean action $A_E$. This, in turn, holds for large values of $M$. Hence, let us look at the turning points in the leading order in $M^{-1}$, which are located at,
\begin{align}
H\eta_{\pm}\approx\frac{k_{\perp}^2+k_z(k_z-qE_0H^{-1})}{k_{\perp}^2+(k_z-qE_{0}H^{-1})^2}\pm i\frac{M}{\sqrt{k_{\perp}^2+(k_z-qE_{0}H^{-1})^2}}
\end{align}
The number of particles is related to the \textit{imaginary} action evaluated for the closed classical trajectory that starts at $\eta_{-}$ and comes back to that point through $\eta_{+}$. The following parametrization turns out  to be a convenient choice for describing this trajectory:
\begin{align}
H\eta(\theta)=\frac{k_{\perp}^2+k_z(k_z-qE_0H^{-1})}{k_{\perp}^2+(k_z-qE_{0}H^{-1})^2}+\frac{iM}{\sqrt{k_{\perp}^2+(k_z-qE_{0}H^{-1})^2}}\sin\theta;\qquad\theta\in\left(-\frac{\pi}{2},\frac{3\pi}{2}\right)~.
\end{align}
Substituting in \ref{actionconstEds}, we can evaluate the action to get
\begin{align}
\mathcal{A}=\frac{M^2}{\sqrt{k_{\perp}^2+(k_z-qE_{0}H^{-1})^2}}\int_{-\pi/2}^{3\pi/2}\frac{\cos^2\theta d\theta}{-iA+B \sin\theta}
\end{align}
where,
\begin{align}
A=\frac{qE_0(k_z-qE_{0}H^{-1})}{k_{\perp}^2+(k_z-qE_{0}H^{-1})^2}&&B=\frac{MH}{\sqrt{k_{\perp}^2+(k_z-qE_{0}H^{-1})^2}}
\end{align}
Then, the Euclidean action can be evaluated to get,
\begin{align}
\mathcal{A}_{E}=-i\mathcal{A}&=\frac{M^2}{\sqrt{k_{\perp}^2+(k_z-qE_{0}H^{-1})^2}}\left[\frac{2\pi\left(-A+\sqrt{A^2+B^2}\right)}{B^2}\right]
\nonumber
\\
&=2\pi\frac{M}{H}-2\pi\frac{qE_0}{H^2}\frac{(k_z-qE_0H^{-1})}{\sqrt{k_{\perp}^2+(k_z-qE_{0}H^{-1})^2}}+\mathcal{O}(M^{-1})
\end{align}
The number of particles is then given by
\begin{align}\label{Euclidean}
n_{\mathbf{k}}\approx \exp \left[-2\pi\left\{\frac{M}{H}-\frac{qE_0}{H^2}\frac{(k_z-qE_0H^{-1})}{\sqrt{k_{\perp}^2+(k_z-qE_{0}H^{-1})^2}}\right\}\right]
\end{align}
The $qE_{0}\rightarrow 0$ limit of the above expression can be easily verified to be consistent our discussions so far on particle production in pure de Sitter, for instance \ref{dsparticle}.
\subsection{Using mode functions}

Finally, we will use the appropriate mode functions to calculate the exact expression for the particle number from the Bogoliubov coefficients. We have deliberately discussed approximate methods first to emphasize the elegance and applicability of these approaches to cases when explicit calculations are impossible. Towards the end of this section we have explicitly verified that the exact expression, in fact, reduces to the results derived in \ref{landaudssec} and \ref{euclideandssec} in the appropriate limits.

The differential equation satisfied by the Fourier transform $\Phi_{\bf k}$, in an expanding universe, in the presence of an electric field,  is given by
	\begin{align}\label{ndsaction}
	\frac{d}{d\eta}\left[a^2\frac{d\Phi_{\mathbf{k}}}{d\eta}\right]+\left\{(\mathbf{k}-q\mathbf{A})^2+m^2a^2\right\}\Phi_{\mathbf{k}}
	\end{align}
	For constant electric field in de Sitter, we obtain
	\begin{align}
	\frac{d^2\Phi_{\mathbf{k}}}{d\eta^2}+\frac{2H}{1-H\eta}\frac{d\Phi_{\mathbf{k}}}{d\eta}+\left\{k_{\perp}^2+\frac{m^2}{(1-H\eta)^2}+\left(k_z+\frac{qE_0\eta}{1-H\eta}\right)^2\right\}\Phi_{\mathbf{k}}=0
	\end{align}
	Let $\Phi_{\mathbf{k}}=(1-H\eta)\psi_{\mathbf{k}}$, so that $\psi_{\mathbf{k}}$ satisfies
	\begin{align}\label{difeqforpsi}
	\psi_{\mathbf{k}}''+\left[k^2+\frac{2qE_0k_z\eta}{1-H\eta}+\frac{m^2-2H^2+q^2E_0^2\eta^2}{(1-H\eta)^2}\right]\psi_{\mathbf{k}}=0
	\end{align}
	We can easily verify that this equation has the correct limits. When $H\rightarrow 0$, we obtain
	\begin{align}
	\psi_{\mathbf{k}}''+\left[k^2+2qE_0k_z\eta+m^2+q^2E_0^2\eta^2\right]\psi_{\mathbf{k}}=0\,;\qquad(H\rightarrow 0),
	\end{align}
	which matches with the time dependent frequency of the Fourier mode of a complex scalar field in a constant electric field in flat spacetime, in the time dependent gauge. On the other hand for $qE_0\rightarrow 0$, we have
	\begin{align}
	\psi_{\mathbf{k}}''+\left[k^2+\frac{m^2-2H^2}{(1-H\eta)^2}\right]\psi_{\mathbf{k}}=0\,\,;\,\,(qE_0\rightarrow 0)
	\end{align}
	which is perfectly consistent with \ref{eominds}. Let us introduce a new variable $z=2ik\eta$ so that \ref{difeqforpsi} simplifies to
\begin{align}\label{diffinWhitform2}
\frac{d^2\psi_{\mathbf{k}}}{dz^2}+\left(-\frac{1}{4}+\frac{\xi}{z}+\frac{\frac{1}{4}-\nu^2}{z^2}\right)\psi_{\mathbf{k}}=0,
\end{align}
where,
\begin{align}\label{definitions}
\xi=i\frac{(k_z-qE_0/H)}{\sqrt{k_{\perp}^2+(k_z-qE_0/H)^2}}\left(\frac{qE_0}{H^2}\right),&&\nu=\sqrt{\frac{9}{4}-\frac{m^2}{H^2}-\frac{q^2E_0^2}{H^4}}
\end{align}
The general solution to this equation can be written in terms of the Whittaker functions as
\begin{align}
\psi_{\mathbf{k}}=C_1\,\, W_{\xi,\nu}\left[2i\sqrt{k_{\perp}^2+\left(k_z-\frac{qE_0}{H}\right)^2}\left(\eta-\frac{1}{H}\right)\right]+C_2\,\, M_{\xi,\nu}\left[2i\sqrt{k_{\perp}^2+\left(k_z-\frac{qE_0}{H}\right)^2}\left(\eta-\frac{1}{H}\right)\right]
\end{align}
From the asymptotic expansion of the Whittaker functions, we get
\begin{align}
W_{\xi,\nu}\left[2i\sqrt{k_{\perp}^2+\left(k_z-qE_0/H\right)^2}\left(\eta-\frac{1}{H}\right)\right]&\approx(H\eta)^\xi\,\exp{\left(-i\sqrt{k_{\perp}^2+\left(k_z-qE_0/H\right)^2}~\eta\right)};~~\textrm{for}~~\eta\rightarrow-\infty
\\
M_{\xi,\nu}\left[2i\sqrt{k_{\perp}^2+\left(k_z-qE_0/H\right)^2}\left(\eta-\frac{1}{H}\right)\right]&\approx \left[2i\sqrt{k_{\perp}^2+\left(k_z-qE_0/H\right)^2}\left(\eta-\frac{1}{H}\right)\right]^{\nu+1/2};~~\textrm{for}~~\eta\approx H^{-1}
\nonumber
\\
&\propto e^{-\nu H t}
\end{align}
where the last relation is true for the case when $\nu$ is purely imaginary, i.e., when $\nu=i|\nu|$. This is similar to the situation in pure de Sitter discussed earlier. Hence, for $\nu=i|\nu|$, we have $M_{\xi,\nu}$ defining the late time vacuum (in terms of cosmic time $t$) and $W_{\xi,\nu}$ defining the in-vacuum (in terms of the conformal time $\eta$). 

Some comments regarding the nature of in-vacuum state are appropriate at this stage. In the $\eta\rightarrow -\infty$ limit, which is the appropriate initial Cauchy slice for the de Sitter spacetime, the modes used to define  the in-vacuum state behave as $\eta ^{\xi}\exp(-ik'\eta)$, where $\xi$ has been defined  in \ref{definitions}. So this state is similar to the Bunch-Davies vacuum, with two crucial differences. (i) First, wave number appearing in the exponential is not the same as that of the Fourier mode $\Phi_{\bf k}$, as in the de Sitter spacetime, rather is modified to $k'=\sqrt{k_{\perp}^2+(k_z-qE_0)^2}$. This modification is  due to the appearance of canonical momenta $k_{i}$, through the combination $k_{i}-qA_{i}$. Since, at the asymptotic past (when $\eta \rightarrow -\infty$), the vector potential becomes $qA_{z}=qE_{0}/H$  it is the combination $k_{\perp}^{2}+(k_{z}-qE_{0}/H)^{2}$ that appears in the solution of the associated mode function. (ii) Second, the pre-factor depends on $\eta$ through a term $\sim \eta ^{\xi}$, where $\xi$ is proportional to the electric field. The presence of this prefactor  $\eta ^{\xi}$ can be understood using the WKB limit. In this limit,   the mode function (in the $\eta \rightarrow -\infty$ limit) takes the form $\exp(i\int dz \sqrt{(1/4)-(\xi/z)})$. Since $z$ is very large one can expand it to the leading order, which upon integration yields a term $\exp(-\xi \ln z)$. This leads to the $\eta^{\xi}$ term in the mode function. Defining the in-vacuum in terms of the exponential part of the mode functions, one immediately observes that these modes indeed carry positive energy alike the Bunch-Davies vacuum state, but with a modified wavenumber $k'$.

The problem of particle production is mathematically identical to our discussion in \ref{singularfield}. In particular, an argument similar to the one used to derive \ref{nforsingular} can employed here to arrive at the following expression for the number of particles:
	\begin{align}\label{dsSchwingerpart}
	n_{\mathbf{k}}=\frac{\cosh\left(\pi |\nu|-\frac{\pi qE_0}{H^2}\frac{(k_z-qE_0/H)}{\sqrt{k_{\perp}^2+(k_z-qE_0/H)^2}}\right)}{e^{2\pi|\nu|}\cosh\left(\pi |\nu|+\frac{\pi qE_0}{H^2}\frac{(k_z-qE_0/H)}{\sqrt{k_{\perp}^2+(k_z-qE_0/H)^2}}\right)-\cosh\left(\pi |\nu|-\frac{\pi qE_0}{H^2}\frac{(k_z-qE_0/H)}{\sqrt{k_{\perp}^2+(k_z-qE_0/H)^2}}\right)}
	\end{align}
	It can be easily verified that we get the correct limiting forms. For, $qE_0\rightarrow 0$, we have
	\begin{align}
	n_{\mathbf{k}}=\frac{1}{e^{2\pi|\nu|}-1},
	\end{align}
	which matches with, say, \ref{tp2}.	On the other hand, if we demand that $M\gg H$, then, \ref{dsSchwingerpart} to the leading order is given by
	\begin{align}\label{Euclidean2}
	n_{\mathbf{k}}\approx e^{-2\pi\left[\frac{ M}{H}-\frac{qE_0}{H^2}\frac{(k_z-qE_0H^{-1})}{\sqrt{k_{\perp}^2+(k_z-qEH^{-1})^2}}\right]}
	\end{align}
    This is also in perfect agreement with \ref{Euclidean}.	Finally, the $H\rightarrow 0$ limit reduces to
	\begin{align}
	n_{\mathbf{k}}=\exp{\left[-\frac{\pi(m^2+k_{\perp}^2)}{qE_0}\right]}
	\end{align}
	which is the correct result for Schwinger effect.

It is worth mentioning that in most of the previous literature, the scale factor as well as the vector potential has been chosen in such form that they diverge in the $H\rightarrow 0$ limit \cite{schwingerinds1,schwingerinds3,schwingerinds2}. This makes the interpretation of the particle number in the limit of vanishing Hubble constant  problematic. This is primarily due to the fact that, the scale factor and gauge choice for the vector potential did not have the appropriate limiting behaviour. Keeping this in mind, in this paper, we have worked with expressions for the scale factor and vector potential which have appropriate limiting behaviour. Then the particle number as well, naturally, leads to the desired expressions for pure de Sitter and pure Schwinger effect, in the $qE_{0}\rightarrow 0$ and $H\rightarrow 0$ limits, respectively.  (For a different view on arriving at the appropriate limits, see \cite{amit}).  

As remarked in the beginning of this subsection, we cannot analytically solve for the mode functions for the most general, time dependent, homogeneous electric field configuration in de Sitter space time. In such cases, one plausible strategy is to employ numerical techniques. However, the approximate methods discussed in \ref{landaudssec} and \ref{euclideandssec} give us an elegant analytic handle. In the following section we will be using one of these approaches, namely the Landau procedure, to study the generalized Schwinger effect in a de Sitter background. 
\section{An example of time-dependent Electric Field in de Sitter}\label{sec2}

In the previous section we have determined the particle production of a complex scalar field in a de Sitter background in presence of a constant electric field. However, in practical situations the electric field is often not a constant but depends on time. Keeping this in mind, we would like to understand particle production due to a time dependent electric field in de Sitter, which may also provide us some insight into the  non-analytic versus analytic behaviour  of the same. 

We start by considering a homogeneous electric in the de Sitter background, satisfying the following condition,
\begin{align}
F^{\mu\nu}F_{\mu\nu}=E^2(\omega_0\eta)
\end{align}
where, the raising and lowering of indices has been performed using the conformally flat form of the metric ansatz, given by \ref{metricds}. Assuming, without any loss of generality, that the electric field is in the $z$-direction the above equation provides us $F_{0z}=E(\omega_0\eta)a(\eta)^{2}$ to be the only non-vanishing component. Given the field tensor, the differential equation governing the vector potential can then be expressed as,
\begin{align}
\frac{d A_{z}}{d\eta}&=-\frac{E(\omega_0\eta)}{(1-H\eta)^2}.
\end{align}
Determination of the vector potential from the above differential equation requires an integration and that requires an explicit expression for the time dependence of electric field. It also requires an additional condition, namely the vector potential should be finite in the $H\rightarrow 0$ limit. To see what this second condition means, let us consider a power law electric field, such that, $E(\omega_{0}\eta)\sim E_{0}(\omega _{0}\eta)^{-s}$, then the vector potential becomes,
\begin{align}
A_{z}=-\frac{E_{0}}{\omega_{0}^{s}}\int \frac{d\eta}{\eta ^{s}(1-H\eta)^{2}}
=-\frac{E_{0}}{\omega_{0}^{s}}\frac{1}{\eta^{s}}\left(1+\frac{1}{H\eta-1}\right)^{s}\frac{_{2}F_{1}(s,1+s,2+s,\frac{1}{1-H\eta})}{(1+s)H(-1+H\eta)}
+\textrm{constant},
\end{align}
so that we obtain,
\begin{align}
\lim _{H\rightarrow 0}A_{z}
=-\frac{E_{0}}{\omega_{0}^{s}}\frac{1}{\eta^{s}}\left(H\eta\right)^{s}\frac{_{2}F_{1}(s,1+s,2+s,1)}{(1+s)H}+\textrm{constant}.
\sim -\frac{E_{0}}{\omega_{0}^{s}}H^{s-1}+\textrm{constant}
\end{align}
Thus, for $s\geq 1$, the vector potential is always finite in the $H\rightarrow 0$ limit and we can choose the constant to be vanishing. While for $s\leq0$ one must take the constant to be $(E_{0}/\omega_{0})H^{s-1}$ to make the vector potential finite in the $H\rightarrow 0$ limit. 

In what follows we will concentrate on electric field of the form $E(\omega _{0}\eta)=E_{0}\{1+f(\omega_{0}\eta)\}$, where $f(\omega_{0}\eta)$ is some arbitrary function which decays for large $\eta$. That is, the electric field becomes a constant at late times. The corresponding  vector potential, having finite $H\rightarrow 0$ limit can be written as
\begin{align}
A_z=-\frac{E_{0}\eta}{1-H\eta}-\frac{E_{0}}{\omega_0}F(\omega_{0}\eta;H)
\end{align}
where the  function $F(\omega_{0}\eta;H)$ satisfies the following differential equation,
\begin{align}
\frac{dF(s)}{ds}=\frac{f(s)}{\left(1-\frac{H}{\omega_{0}}s\right)^2}
\end{align}
It is, of course, convenient to work with $F$ rather than $f$, which is what we will do.

A complex massive scalar field, in the background  of the time dependent electric field in de Sitter universe, will have Fourier modes which again satisfy the equation for a time dependent harmonic oscillator. In this case, the oscillator associated with $k$-th wave mode will have unit mass and a time dependent frequency given by
\begin{align}\label{omegaeta}
\omega_{k}^2(\eta)&=k_{\perp}^2+\frac{m^2}{(1-H\eta)^2}+\left(k_{z}-\frac{qE_{0}\eta}{1-H\eta}-\frac{qE_{0}}{\omega_0}F(\omega_{0}\eta;H)\right)^{2}
\end{align}
Here, $k_{\perp}^{2}=k^{2}-k_{z}^{2}$ is the wave vector component transverse to the direction of electric field. One cannot, of course, solve for the mode functions for arbitrary $F$. To illustrate the use of Landau procedure we shall confine ourselves to a specific choice, viz. $f(\omega_{0}\eta)=2f_{2}(\omega_{0}\eta)^{-3}\{1-2H\eta\}\{1-H\eta\}^{-1}$, where $f_2$ is a constant. With this kind of electric field, the contribution to the vector potential becomes $-f_{2}(\omega_{0}\eta)^{-2}(1-H\eta)^{-2}$. Thus the time dependent frequency in the large $\eta$ limit, but with small $H\eta$, can be expanded as, 
\begin{align}\label{omegaeta}
\omega_{k}^2(\eta)&\approx k_{\perp}^2+m^{2}\left(1+2H\eta+3H^{2}\eta^{2}\right)+\left(qE_{0}\eta\left(1+H\eta+H^{2}\eta^{2}\right)
+\frac{qE_{0}f_{2}}{\omega_0^{3}\eta^{2}}\left(1+2H\eta+3H^{2}\eta^{2}\right)\right)^{2}
\nonumber
\\
&\approx\left(q^{2}E_{0}^{2}+3m^{2}H^{2}\right)\eta^{2}+\left(2m^{2}H+2qE_{0}\frac{3H^{2}f_{2}qE_{0}}{\omega_{0}^{3}} \right)\eta 
+\left(k_{\perp}^{2}+m^{2}\right)+\left(\frac{3H^{2}f_{2}qE_{0}}{\omega_{0}^{3}}\right)^{2}
\nonumber
\\
&\equiv A\eta^{2}+B\eta+C
\end{align}
Here, the last relation defines the  constants $A$, $B$ and $C$ in terms of the parameters appearing in this model, e.g., the electric field strength $E_{0}$, Hubble constant $H$, the inverse time scale $\omega _{0}$ etc. The corresponding expansion for $\omega_k$ is: 
\begin{align}\label{omegaeta}
\omega _{k}&=\sqrt{A}\eta\left(1+\frac{B}{A\eta}+\frac{C}{A\eta^{2}}\right)^{1/2}
\nonumber
\\
&=\sqrt{A}\eta\left(1+\frac{B}{2A\eta}+\frac{C}{2A\eta^{2}}-\frac{1}{8}\frac{B^{2}}{A^{2}\eta ^{2}}\right)
\nonumber
\\
&=\sqrt{A}\eta+\frac{B}{2\sqrt{A}}+\frac{C}{2\sqrt{A}\eta}-\frac{1}{8}\frac{B^{2}}{A^{3/2}\eta}
\end{align}
Having derived this expression, one can invoke the Landau procedure to extract non-analytic part of the particle number. This requires one to analytically continue the range of $\eta$ from $\{-\infty,(1/H)\}$ to $\{-\infty,\infty\}$. Further, using the WKB method, one can determine the in-states and out-states associated with the Fourier modes at $\eta=\mp \infty$ respectively. The Bogoliubov coefficient connecting them can be obtained by treating $\eta$ as a complex variable and rotating it in the complex plane from $\textrm{Arg}[\eta]=0$ to $\textrm{Arg}[\eta]=\pi$. This provides the non-analytic part of the particle number to be dependent on the coefficient of $(1/\eta)$ in the expression for $\omega_{k}$, which reads,
\begin{align}\label{omegaeta}
n_{k}&=\exp \left[-2\pi\left(\frac{C}{2\sqrt{A}}-\frac{1}{8}\frac{B^{2}}{A^{3/2}}\right) \right]
\nonumber
\\
&=\exp \left[-2\pi\left(\frac{\left(k_{\perp}^{2}+m^{2}\right)+\left(\frac{3H^{2}f_{2}qE_{0}}{\omega_{0}^{2}}\right)^{2}}{2\sqrt{\left(q^{2}E_{0}^{2}+3m^{2}H^{2}\right)}}-\frac{1}{8}\frac{\left(2m^{2}H+2qE_{0}\frac{3H^{2}f_{2}qE_{0}}{\omega_{0}^{2}} \right)^{2}}{\left(q^{2}E_{0}^{2}+3m^{2}H^{2}\right)^{3/2}}\right) \right]
\end{align}
Note that in the $H\rightarrow 0$ limit the particle number becomes $\exp[-(k_{\perp}^{2}+m^{2})/qE_{0}]$ irrespective of presence of $f_{2}$. This is what we expect, as the Landau procedure picks up the non-analytic part which is given by the coefficient of constant term irrespective of other terms in the expansion. This assures that Landau procedure works in de Sitter spacetime as well and yields the non-analytic part of the particle number for time dependent electric fields in de Sitter, while remaining compatible with flat spacetime limit. 

However, the particle number presented above do not yield the de Sitter particle production  as the electric field vanishes. This is due to the fact that, for Landau procedure to work we have analytically extended the de Sitter spacetime to cover the full range of $\eta$, namely $\eta \in (-\infty,\infty)$ and hence the background spacetime is not exactly the de Sitter background we want to   work with. Besides, this feature is also present in the context of constant electric field, as evident from \ref{omegaforHzero}. This suggest that even though Landau procedure is an useful method to understand non-analytic behaviour of the particle production in time dependent electric field, it has its limitations when applied in the context of  an expanding universe.
\section{Summary}

The previous sections discussed several aspects of particle production in an expanding universe and its possible correspondence with  the generalized Schwinger effect. Given the fact that both these phenomena have been investigated extensively in the literature, it is useful to highlight the new --- conceptual and technical --- results in this paper.

\begin{itemize}

\item The  correspondence between time dependent electric field and an expanding universe has been noticed earlier,  one of the earliest works  being Ref. \cite{Padmanabhan:1991uk}  and a more recent one being \cite{Martin2007}. However, this correspondence was noticed at a formal level, and was not adequately exploited. In this work we have taken this further and applied this formalism to connect some well known cosmological spacetimes to specific time dependent electric fields and vice versa. 

For example, we studied the cosmological analogue of Sauter type electric field and showed that they  possess non-analytic behaviour as the scale factor approaches that of the radiation dominated universe. Further, through this correspondence we could provide estimation of particle number for non-trivial electric field configuration using our knowledge about the results for the expanding universe. Starting from the de Sitter (or, quasi-de Sitter) spacetime we have determined the corresponding electric field and hence the corresponding particle production. 
We also discovered a time dependent electric field in flat spacetime which can lead, in a specific limit,  to a  Planck spectrum of particles at late times. (This `analogue black hole' model deserves further exploration expecially as regards back reaction. To our knowledge such electric fields have not been explored earlier.) It will also be interesting to take this correspondence further, to the level of two-point functions, and analyze the analogue of inflationary power spectrum in the context of generalized Schwinger effect. 

\item In our earlier work \cite{Rajeev2018} we used an asymptotic expansion of the electric field, and  identified the terms responsible for non-analytic behaviour of the particle number. Through the correspondence between generalized Schwinger effect and \PPEU, we have  determined the corresponding factors responsible for the non-analytic behaviour of particle number in an expanding background. In particular, we have shown that the coefficients of $\eta^{-1}$ in the expansion of $a(\eta)$ as well as $a^{-1}(\eta)$ controls the non-analytic behaviour of the particle number.

\item We have also clarified some of the conceptual issues which arise in the study of particle production in the de Sitter spacetime. The standard procedure of computing the   Bogoliubov coefficients associated with mode functions, does not work satisfactorily  if the mass $M$ of the scalar field is such that $(M/H)<(3/2)$ (which includes the massless case). We have  addressed this issue using the transformation of a time dependent harmonic oscillator to that of a constant frequency oscillator. This approach provides a conceptual basis for computing  the particle number for \emph{all} values of $M/H$, including the massless case. Interestingly, the massless limit involves certain subtleties, and the asymptotic limits depend on the order in which the limits are taken. The number of particle produced remains finite throughout the expansion history, and diverges only when $a\to\infty$. For $(M/H)>(3/2)$, the particle number  remains finite asymptotically and matches with the mode function analysis under appropriate limits. 

\item In the last part, we discussed the case of a constant electric field in de Sitter spacetime, using three different approaches and compared the results.  First, we have described how the Landau procedure can be used to infer the non-analytic part of the particle number and we have shown that it reproduces the correct  result. Second we use  the Euclidean action approach to obtain the asymptotic limit of the same result. Finally, we have studied this case using the conventional approach based on mode functions. In all the cases, we  have worked in a gauge which allows taking appropriate limits and we explicitly verify these limits. (This has been an issue in some of the previous works in the literature.)

\item Taking a cue from this discussion and our earlier results  in \cite{Rajeev2018}, we describe how one may go about studying particle production due to a time dependent electric field in de Sitter. Using the technique due to Landau, we have been able to obtain the non-analytic part of the particle production in the context of a specific time dependent field in de Sitter. Even though we could retrieve the desired Schwinger result in appropriate limit, the general structure of the particle number is more complicated and deserves further attention. We hope to study this in a future work.

\end{itemize}

\section*{Acknowledgement}

KR is supported by Senior Research Fellowship of the Council of Scientific \& Industrial Research (CSIR), India. Research of SC is supported by INSPIRE Faculty Fellowship (Reg. No. DST/INSPIRE/04/2018/000893) from Department of Science and Technology, Government of India. Research of TP is partially supported by the J.C.Bose Fellowship of Department of Science and Technology, Government of India.
\appendix	
\labelformat{section}{Appendix #1}
\labelformat{subsection}{Appendix #1}
\labelformat{subsubsection}{Appendix #1}
\section{Derivation for radiation dominated universe}\label{radiation}

Let us begin by looking at the $\lambda\rightarrow 0$ limits of $\tilde{\omega}_{\pm}$. We have
\begin{align}
\sqrt{k^2+M^2(A+B+C)}
&=\sqrt{k^2+M^2\left(A+\frac{a_{0}\sqrt{A}}{\lambda}+\frac{a_{0}^{2}}{4\lambda ^{2}}\right)}
\nonumber
\\
&\simeq \frac{Ma_{0}}{2\lambda}\left\{1+4\frac{\sqrt{A}}{a_{0}}\lambda + 4\frac{k^2+M^2 A}{M^{2}a_{0}^{2}}\lambda ^{2} \right\}^{1/2}
\nonumber
\\
&\simeq \frac{Ma_{0}}{2\lambda}+M\sqrt{A}+\lambda \frac{k^{2}}{Ma_{0}}+\mathcal{O}(\lambda^2)
\end{align} 
as well as
\begin{align}
\sqrt{k^2+M^2(A-B+C)}
=\frac{Ma_{0}}{2\lambda}-M^2\sqrt{A}+\lambda \frac{k^{2}}{Ma_{0}}+\mathcal{O}(\lambda^2)
\end{align} 
Therefore, the following limits of the characteristic frequencies $\omega _{\pm}$ are obtained,
\begin{align}
\omega _{+}=\frac{Ma_{0}}{2\lambda}+\lambda \frac{k^{2}}{Ma_{0}}+\mathcal{O}(\lambda^2);
\qquad
\omega _{-}=M\sqrt{A}+\mathcal{O}(\lambda^2)
\end{align} 
Thus the $\lambda\rightarrow 0$ limit of \ref{particlenumber}, with the parameters $B$ and $C$ as given in \ref{parameter}, becomes
\begin{align}
\lim _{\lambda \rightarrow 0}n_{\mathbf{k}}&=\frac{\cosh \left(\frac{2\pi M\sqrt{A}}{\lambda}\right)+\cosh \left(\frac{2\pi Ma_{0}}{2\lambda ^{2}}\right)}{\cosh \left(\frac{2\pi Ma_{0}}{2\lambda ^{2}}+\frac{2\pi k^{2}}{Ma_{0}}\right)-\cosh \left(\frac{2\pi M\sqrt{A}}{\lambda}\right)}
=\exp\left(\frac{-2\pi k^2}{Ma_0^2}\right)
\end{align}

\section{Discontinuity in asymptotic behaviour of $n_{\mathbf{k}}$ at $\nu=3/2$}\label{disc}

Let us first consider the particle number as $\nu=3/2$.  From \ref{particlemassless}, we get
\begin{align}
\lim_{3/2}\mathcal{N}_{\mathbf{k}}(\nu,a)=\frac{H^2}{4k^2a}
\end{align}
One can see that \ref{nu_then_a} easily follows. For the second limit, let us consider the large $a(\eta)$ limit of $n_{k}$ for a fixed value of $\nu$. We start with the series expansion for $J_{\nu}$.
\begin{align}
J_{\nu}(x)=x^{\nu}\sum_{l=0}^{\infty}\frac{(-1)^l}{l!\Gamma(\nu+l+1)}x^{2l}
\end{align}
The Hankel function can be written in terms of the $J_{\nu}$ as
\begin{align}
H_{\nu}^{(1)}(x)&=\frac{J_{-\nu}(x)-e^{-i\pi\nu}J_{-\nu}(x)}{i\sin(\pi\nu)}\\
&=\left(\frac{1}{i\sin(\pi\nu)}\right)\left[x^{\nu}\sum_{l=0}^{\infty}\frac{(-1)^l}{l!\Gamma(\nu+l+1)}x^{2l}-e^{-i\pi\nu}x^{-\nu}\sum_{l=0}^{\infty}\frac{(-1)^l}{l!\Gamma(-\nu+l+1)}x^{2l}\right]
\end{align}
Therefore, from \ref{f_for_general}, the function $f_{\mathbf{k}}$ has the following behaviour at the leading order in $a^{-1}$ as $a\sim\infty$.
\begin{align}
f_\mathbf{k}\approx\frac{2^{\nu -\frac{1}{2}}}{\sqrt{\pi }} a^{\nu -\frac{3}{2}} \Gamma (\nu ) \left(\frac{k}{H}\right)^{\frac{1}{2}-\nu }
\end{align}
The particle number can then be computed from as \ref{betasquared}
\begin{align}
n_{k}\approx\frac{a^{2\nu}}{4f_0^2k}\left[\sqrt{m^2+\frac{k^2}{a^2}}+\left\{(5 \nu -3)\nu+\frac{9}{4}\right\}\frac{H^2}{\sqrt{m^2+\frac{k^2}{a^2}}}\right]\qquad ;\forall\nu>0 
\end{align}
where,
\begin{align}
f_0=\sin(\pi\nu)\Gamma(1-\nu)\left(\frac{k}{2H}\right)^{2\nu}
\end{align}
This implies that
\begin{align}
\lim_{a\rightarrow\infty}\mathcal{N}_{\mathbf{k}}(\nu,a)=\left(\frac{3\sqrt{3}H^4}{2k^4\pi}\right)\left(\nu-\frac{3}{2}\right)^{-\frac{1}{2}}+\mathcal{O}\left((\nu-3/2)^{1/2}\right)
\end{align}
By taking $\nu\rightarrow3/2$ in this equation we arrive at \ref{a_then_nu}.
\section{Constant frequency mapping}\label{constant_freq_mapping}

Let us briefly review quantization of a simple harmonic oscillator $q$ of constant mass $m_0$ and constant frequency $\omega_0$ as a warm up exercise. The operator equation satisfied by $q$ is given by
\begin{align}\label{sho}
q''+\omega_0^2q=0
\end{align}
where `prime' denotes derivative with respect to $\eta$. The general solutions to this equation can be written as
\begin{align}\label{definaadag}
q(\eta)=a  \frac{\xi(\eta)}{\sqrt{2\omega_0}}+a^{\dagger}\frac{\xi^*(\eta)}{\sqrt{2\omega_0}} 
\end{align}
where, 
\begin{align}\label{solsho}
\xi=\frac{e^{-i\omega_0\eta}}{\sqrt{m_0}}
\end{align}
and the operators $a$ and $a^{\dagger}$ to satisfy $[a,a^{\dagger}]=1$. Given \ref{solsho} we see that $\xi$ and $\xi^*$ satisfy the following Wronskian condition,
\begin{align}\label{wronksho}
im_0\left[\xi^*\xi'-\xi(\xi^*)'\right]=\omega_0
\end{align}
Let us now consider a time dependent harmonic oscillator, whose equation of motion given by
\begin{align}\label{timedependentho}
\left\{m(\eta)q'\right\}'+\omega^2(\eta)q=0
\end{align}
Clearly, the solutions to this equation are not simple phases as in the case of a simple harmonic oscillator. Let us denote a solution to \ref{timedependentho} by $\psi$. Then it follows that, 
\begin{align}\label{wronskt}
\frac{d}{d\eta}\left\{m(\eta)\left[\psi^*\psi'-\psi(\psi^*)'\right]\right\}=0
\end{align}
Thus motivated by \ref{wronksho}, we demand that the following Wronskian condition to hold,
\begin{align}\label{wronsktdho}
im(\eta)\left[\psi^*\psi'-\psi(\psi^*)'\right]=\omega_i
\end{align}
where $\omega_i\equiv\omega(\eta_i)$, for some choice of $\eta_i$. Without any loss of generality, one may write $\psi$ as $\psi=f(\eta)e^{-i\theta(\eta)}$, where $f(\eta)$ is real. Then \ref{wronsktdho} simplifies to
\begin{align}
\frac{d\theta}{d\eta}=\frac{\omega_i}{m(\eta)f^2}
\end{align}
while, from \ref{timedependentho} we get the differential equation satisfied by $f$ to be
\begin{align}
\left\{m(\eta)f'\right\}'+\omega^2(\eta)f=\frac{\omega_i^2}{m(\eta)f^3}
\end{align}
This motivates us to introduce a new time coordinate $\tau$ such that,
\begin{align}
d\tau=\frac{d\eta}{mf^2}
\end{align}
so that $d\theta=\omega_id\tau$ and
\begin{align}
\tilde{\psi}\equiv\frac{\psi}{f}=e^{-i\omega_i\tau}
\end{align}
Comparing this with \ref{solsho}, we may interpret $\tilde{\psi}$ as a solution to a time independent harmonic oscillator $Q=q/f$ of unit mass and frequency $\omega_i$. The equation of motion of $Q$ is given by
\begin{align}
\frac{d^2Q}{d\tau^2}+\omega_i^2Q=0
\end{align}

Let us now derive the same starting from the Schr\"{o}dinger picture. Note that in the Schr\"{o}dinger picture the state ket changes with time, but the operators do not. This exercise will provide an internal consistency of these results. Let $\Psi(Q,\tau)$ be the Schr\"{o}dinger wavefunction for the state $\ket{0}$ in terms of $Q$. By definition we have
\begin{align}
\Psi(Q;\tau)= \left(\frac{\omega_i}{\pi}\right)^{1/4}\exp\left(-\frac{\omega_i Q^2}{2}-\frac{i}{2}\omega_i\tau\right)
\end{align}
On the other hand, the Schr\"{o}dinger wavefunction $\Phi(q,\eta)$ for $\ket{0}$ in terms of $q$ is given by (see \cite{Rajeev:2017uwk} for details)
\begin{align}
\Phi(q;\eta)&=\frac{1}{\sqrt{f}}e^{i\left(m\frac{f'}{2f}q^2\right)}\Psi\left(Q=\frac{q}{f};\tau(\eta)\right)\\
&=\left(\frac{\omega_i}{f^2\pi}\right)^{1/4}\exp\left[-\left(\frac{\omega_i}{2f^2}-im\frac{f'}{2f}\right)q^2-\frac{i}{2}\omega_i\tau(\eta)\right]
\end{align}
The expectation value of the instantaneous Hamiltonian of the $q$-system in this state is given by
\begin{align}\label{averageH}
\int_{-\infty}^{\infty}dq\,\,\Phi^*(q;\eta)\left[-\frac{1}{2m}\partial^2_{q}+\frac{m\omega^2}{2}q^2\right]\Phi(q;\eta)&=\frac{m \left[f^2 \omega ^2+(f')^2\right]}{4 \omega_i}+\frac{\omega_i}{4 f^2 m}\\
&=\omega\left\{\frac{mf^{2}}{4(\omega\omega_i)}\left[\left(-\omega+\frac{\omega_i}{mf^{2}}\right)^2+\left(\frac{f'}{f}\right)^2\right]+\frac{1}{2}\right\}
\end{align}
Comparing the right hand side with $\omega\{\bar{n}(\eta)+1/2\}$ we see that 
\begin{align}
\bar{n}(\eta)=\frac{mf^{2}}{4(\omega\omega_i)}\left[\left(-\omega+\frac{\omega_i}{mf^{2}}\right)^2+\left(\frac{f'}{f}\right)^2\right]
\end{align}
which is the result quoted in \ref{averagen}. 
\bibliography{Schwinger_in_ds}

\bibliographystyle{utphys1}

\end{document}